\documentclass[nonacm]{acmart}

\AtBeginDocument{%
  }

\usepackage[english,main=english]{babel}
\usepackage{tikz}
\usepackage{forest}

\definecolor{carminepink}{rgb}{0.92, 0.3, 0.26}
\definecolor{celadon}{rgb}{0.67, 0.88, 0.69}
\definecolor{harvestgold}{rgb}{0.85, 0.57, 0.0}
\definecolor{majorblue}{rgb}{0.25, 0.41, 0.88}
\definecolor{orange}{rgb}{1.0, 0.5, 0.0}

\usepackage{tikz}
\usetikzlibrary{mindmap,shapes, positioning}
\usepackage{smartdiagram}
\usesmartdiagramlibrary{additions}
\usepackage{forest}
\usetikzlibrary{shadows}
\usepackage{subfigure}
\usetikzlibrary{mindmap}
\usepackage{siunitx}
\usepackage{etoolbox}
\usepackage{nicematrix}
\usepackage{pifont}
\usepackage{enumitem}
\usepackage{fontawesome5}
\usepackage{graphicx}
\usepackage{ragged2e} 
\usepackage{makecell} 
\usepackage{diagbox} 
\usepackage[T1]{fontenc} 
\usepackage[utf8]{inputenc}

\usepackage{tabularray}
\usepackage{adjustbox}
\usepackage{colortbl}
\usetikzlibrary{positioning}
\usepackage{tablefootnote}
\usepackage{amsmath,amsfonts}
\usepackage{algorithmic}
\usepackage{algorithm}
\usepackage{array}
\usepackage{textcomp}
\usepackage{stfloats}
\usepackage{url}
\usepackage{booktabs} 
\usepackage{multirow}
\usepackage{tabularx}
\usepackage{hyperref}
\usepackage{cleveref}
\usepackage{multicol}
\usepackage{subcaption}
\DeclareUnicodeCharacter{02BC}{'}
\usepackage{indentfirst}
\usepackage{tikz}
\usetikzlibrary{shapes.geometric, arrows}
\tikzstyle{box} = [rectangle, rounded corners, minimum width=3cm, minimum height=1cm, text centered, 
]
\tikzstyle{leaf} = [rectangle, minimum width=4cm, minimum height=1cm, text centered, 
fill=green!10]
\tikzstyle{line} = [draw, -latex']

\usepackage{longtable}
\usepackage{CJKutf8}
\usepackage[utf8]{inputenc}

\definecolor{lightcoral}{rgb}{0.94, 0.5, 0.5}
\definecolor{lightgreen}{rgb}{0.56, 0.93, 0.56}
\definecolor{harvestgold}{rgb}{0.85, 0.57, 0.0}
\definecolor{brightlavender}{rgb}{0.75, 0.58, 0.89}
\definecolor{capri}{rgb}{0.0, 0.75, 1.0}
\definecolor{carminepink}{rgb}{0.92, 0.3, 0.26}
\definecolor{celadon}{rgb}{0.67, 0.88, 0.69}
\definecolor{darkpastelgreen}{rgb}{0.01, 0.75, 0.24}
\definecolor{DeepSkyBlue4}{RGB}{0,104,139}
\definecolor{DeepPurple}{RGB}{48,0,96}
\definecolor{DeepGreen}{RGB}{0,102,0}

\definecolor{softblue}{RGB}{100,149,237}
\definecolor{softgreen}{RGB}{144,238,144}
\definecolor{softpurple}{RGB}{230,190,255}
\definecolor{softorange}{RGB}{255,160,122}
\definecolor{softpink}{RGB}{255,182,193}
\definecolor{majorblue}{RGB}{175,199,232}
\definecolor{majororange}{RGB}{240,145,72}
\definecolor{majoryellow}{RGB}{255,152,150}
\definecolor{milkyellow}{RGB}{255,255,204}

\definecolor{raspberry}{RGB}{200,111,103}

\usepackage{cellspace}
\begin{document}

\title{Towards Robust and Secure Embodied AI: A Survey on Vulnerabilities and Attacks}

\author{Wenpeng Xing}
\email{wpxing@zju.edu.cn}

\author{Minghao Li}
\email{mhli@s.hlju.edu.cn}

\author{Mohan Li}
\email{limohan@gzhu.edu.cn}


\author{Meng Han}
\authornotemark[1]
\email{mhan@zju.edu.cn}

\renewcommand{\shortauthors}{Xing, et al.}

\begin{abstract}

 Embodied AI systems, including robots and autonomous vehicles, are increasingly integrated into real-world applications, where they encounter a range of vulnerabilities stemming from both environmental and system-level factors. These vulnerabilities manifest through sensor spoofing, adversarial attacks, and failures in task and motion planning, posing significant challenges to robustness and safety. 
Despite the growing body of research, existing reviews rarely focus specifically on the unique safety and security challenges of embodied AI systems. Most prior work either addresses general AI vulnerabilities or focuses on isolated aspects, lacking a dedicated and unified framework tailored to embodied AI. This survey fills this critical gap by: (1) categorizing vulnerabilities specific to embodied AI into exogenous (e.g., physical attacks, cybersecurity threats) and endogenous (e.g., sensor failures, software flaws) origins; (2) systematically analyzing adversarial attack paradigms unique to embodied AI, with a focus on their impact on perception, decision-making, and embodied interaction; (3) investigating attack vectors targeting large vision-language models (LVLMs) and large language models (LLMs) within embodied systems, such as jailbreak attacks and instruction misinterpretation; (4) evaluating robustness challenges in algorithms for embodied perception, decision-making, and task planning; and (5) proposing targeted strategies to enhance the safety and reliability of embodied AI systems.
 By integrating these dimensions, we provide a comprehensive framework for understanding the interplay between vulnerabilities and safety in embodied AI. 

 \end{abstract}

\keywords{large vision language models, large language models, embodied AI, adversarial attack.}

\settopmatter{printacmref=false}
\renewcommand\footnotetextcopyrightpermission[1]{}

\maketitle

\section{Introduction}\label{sec:intro}

\begin{figure*}
    \centering
    \includegraphics[width=\linewidth]{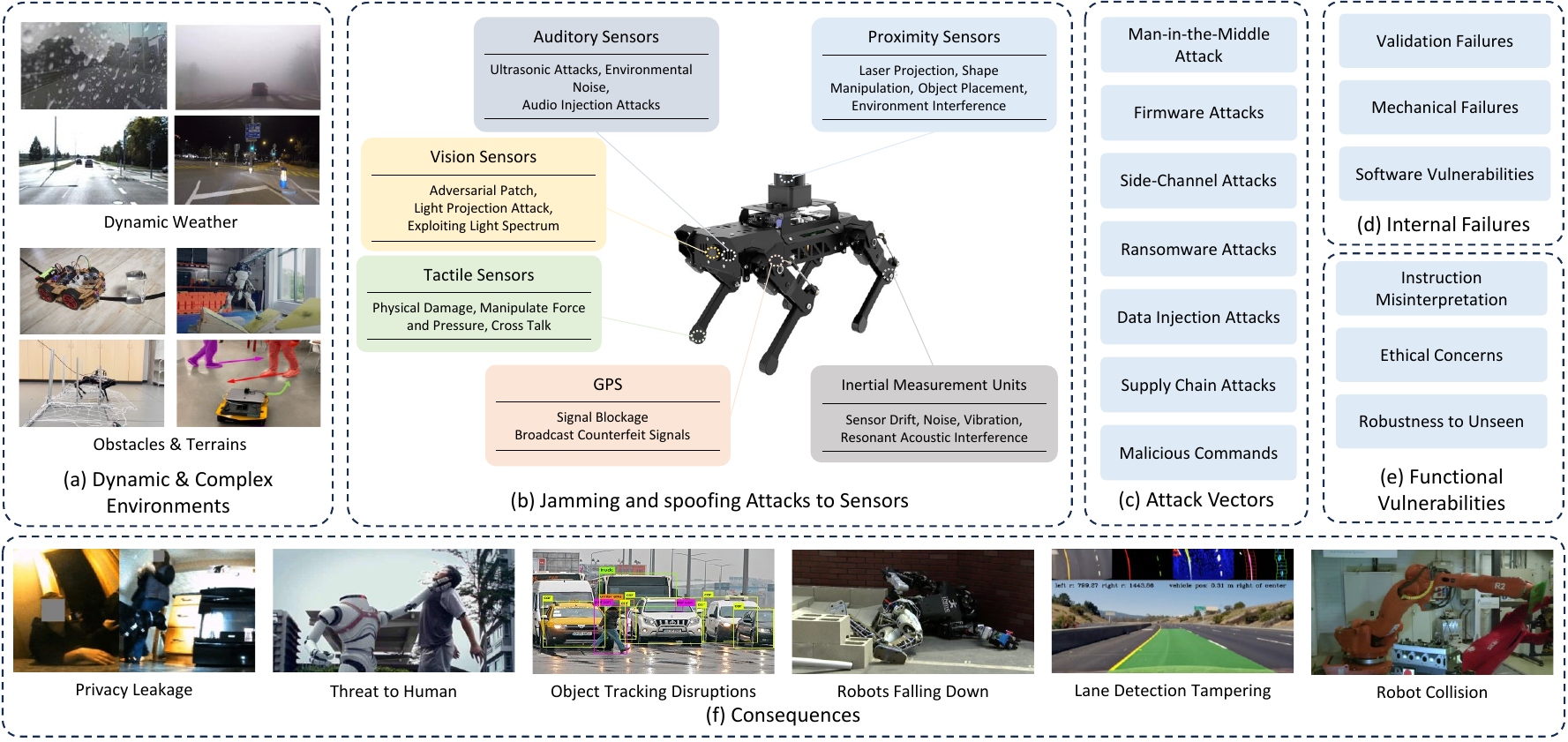}
    \caption{Security Threats and Challenges in Embodied AI Systems.}
    \label{fig:risk_cat}
\end{figure*}

The rapid advancement of AI has established \textit{Embodied AI} as a key technology in domains such as autonomous driving, industrial automation, and smart home systems. By combining perception, decision-making, and actuation, these systems excel in handling complex real-world tasks. However, as shown in Fig. \ref{fig:risk_cat}, their reliance on intricate interactions between sensors, actuators, and algorithms also exposes them to a broad spectrum of vulnerabilities, including dynamic and complex environments, jamming and spoofing attacks on sensors, and system failures. These risks raise concerns over unauthorized actions and reputation damage, underscoring the critical need for robust security measures to ensure their safe and reliable deployment.

We identify three key characteristics of robust embodied systems \cite{cangelosi2018babies}: {autonomy}, {embodiment}, and {cognition}. {Autonomy} refers to the system's capacity to make informed, independent decisions, enabling it to adapt to dynamic and unpredictable scenarios. However, this independence also introduces vulnerabilities, such as decision-making errors in complex environments. {Embodiment} denotes the ability to interact with physical environments, integrating physical presence with decision-making processes to achieve seamless interaction. {Cognition} encompasses the system's capacity for understanding, reasoning, and interpreting its actions, ensuring that its behavior aligns with both internal goals and external constraints. Yet, cognitive processes can also be exploited through sensor-to-model attacks or manipulation of learned models. 

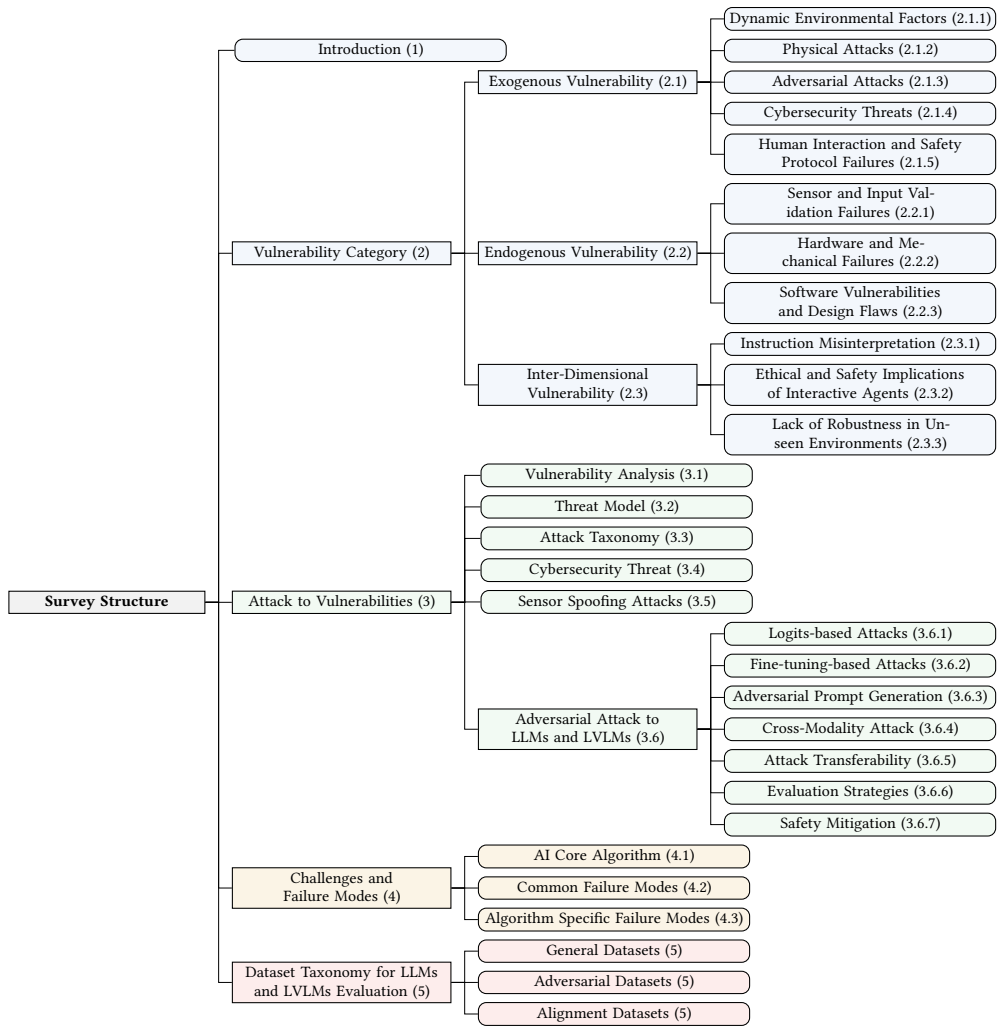
\begin{figure*}
\centering
\tikzset{
        my node/.style={
            draw,
            align=center,
            thin,
            text width=2.8cm, 
            fill=black!5,
            font=\scriptsize
        },
        my leaf/.style={
            draw,
            align=center,
            thin,
            text width=3.5cm, 
            rounded corners=3,
            fill=black!2,
            font=\scriptsize
        }
}
\forestset{
  every leaf node/.style={
    if n children=0{#1}{}
  },
  every tree node/.style={
    if n children=0{minimum width=1em}{#1}
  },
}
\begin{forest}
    nonleaf/.style={font=\scriptsize},
     for tree={%
        every leaf node={my leaf, font=\scriptsize},
        every tree node={my node, font=\scriptsize, l sep-=4.5pt, l-=1.pt},
        anchor=west,
        inner sep=1.5pt,
        l sep=10pt,
        s sep=3pt,
        fit=tight,
        grow'=east,
        edge={ultra thin},
        parent anchor=east,
        child anchor=west,
        if n children=0{}{nonleaf}, 
        edge path={
            \noexpand\path [draw, \forestoption{edge}] (!u.parent anchor) -- +(5pt,0) |- (.child anchor)\forestoption{edge label};
        },
        if={isodd(n_children())}{
            for children={
                if={equal(n,(n_children("!u")+1)/2)}{calign with current}{}
            }
        }{}
    }
    [
  \textbf{Survey Structure}, 
  fill=black!5, text width=2.5cm
        [Introduction (\ref{sec:intro}), fill=majorblue!15
        ]
        [Vulnerability Category (\ref{sec:robot_vul}), fill=majorblue!15
            [Exogenous Vulnerability (\ref{sec:ex_vul}), fill=majorblue!15
                [Dynamic Environmental Factors (\ref{sec:dyn_env_fac}), fill=majorblue!15]
                [Physical Attacks (\ref{sec:cyb_attack}), fill=majorblue!15]
                [Adversarial Attacks (\ref{sec:adv_attack}), fill=majorblue!15]
                [Cybersecurity Threats (\ref{sec:cybersec_threat}), fill=majorblue!15]
                [Human Interaction and Safety Protocol Failures (\ref{sec:human_interc_sef_proto}), fill=majorblue!15]
            ]
            [Endogenous Vulnerability (\ref{sec:end_vul}), fill=majorblue!15
                [Sensor and Input Validation Failures (\ref{sec:sensor_valid_fail}), fill=majorblue!15]
                [Hardware and Mechanical Failures (\ref{sec:hard_mecha_failure}), fill=majorblue!15]
                [Software Vulnerabilities and Design Flaws (\ref{sec:softwa_vul}), fill=majorblue!15]
            ]
            [Inter-Dimensional Vulnerability (\ref{sec:cros_d_vul}), fill=majorblue!15
                [Instruction Misinterpretation (\ref{sec:instruc_misint}), fill=majorblue!15]
                [Ethical and Safety Implications of Interactive Agents (\ref{sec:ethical_interc_ag}), fill=majorblue!15]
                [Lack of Robustness in Unseen Environments (\ref{sec:lack_robost}), fill=majorblue!15]
            ]
        ]
        [Attack to Vulnerabilities (\ref{sec:adv_attack_to_vul}), fill=celadon!15
            [Vulnerability Analysis (\ref{sec:reason_beh_LLM_VLM}), fill=celadon!15]
            [Threat Model (\ref{sec:threat_model}), fill=celadon!15]
            [Attack Taxonomy (\ref{sec:attack_taxo}), fill=celadon!15]
            [Cybersecurity Threat (\ref{sec:syb_sec_attack}), fill=celadon!15]
        [Sensor Spoofing Attacks (\ref{sec:sensor_spoof_attack}), fill=celadon!15]
        [Adversarial Attack to LLMs and LVLMs (\ref{adv_attack_llms}), fill=celadon!15
            [Logits-based Attacks (\ref{sec:logits_based_attack}), fill=celadon!15]
            [Fine-tuning-based Attacks (\ref{sec:fintuning_based_attacks}), fill=celadon!15]
            [Adversarial Prompt Generation (\ref{sec:adv_prompt_gen}), fill=celadon!15]
            [Cross-Modality Attack (\ref{sec:cros——model_attack}), fill=celadon!15]
            [Attack Transferability (\ref{sec:attack_transfer}), fill=celadon!15]
            [Evaluation Strategies (\ref{sec:evaluation_strage}), fill=celadon!15]
            [Safety Mitigation (\ref{sec:safetyu_mitigation}), fill=celadon!15]
        ]
        ]
        [Challenges and Failure Modes (\ref{sec:fail_ai_core}), fill=harvestgold!10
            [AI Core Algorithm (\ref{sec:ai_core_alg}), fill=harvestgold!10]
            [Common Failure Modes (\ref{sec:cfm}), fill=harvestgold!10]
            [Algorithm Specific Failure Modes (\ref{sec:algorithm_specif_fail}), fill=harvestgold!10]
        ]
        [Dataset Taxonomy for LLMs and LVLMs Evaluation (\ref{sec:dataset_taxono_llm_vlm}), fill=carminepink!10
            [General Datasets (\ref{sec:general_dataset}), fill=carminepink!10]
            [Adversarial Datasets (\ref{sec:adv_dataset}), fill=carminepink!10]
            [Alignment Datasets (\ref{sec:align_dataset}), fill=carminepink!10]
        ]
    ]
\end{forest}
\caption{The structure of the survey with section references and limited section width.}
\label{fig:article_structure}
\end{figure*}

The remainder of this article is structured to systematically address the vulnerabilities, attack vectors, and mitigation strategies in embodied AI systems, as illustrated in Fig.~\ref{fig:article_structure}. Section~\ref{sec:robot_vul} introduces a detailed taxonomy of vulnerabilities in embodied AI, categorized into exogenous, endogenous, and inter-dimensional risks, with further exploration of specific issues such as physical attacks (\ref{sec:cyb_attack}), sensor validation failures (\ref{sec:sensor_valid_fail}), and ethical challenges in interactive agents (\ref{sec:ethical_interc_ag}). Section~\ref{sec:reason_beh_LLM_VLM} examines the relationship between vulnerabilities and attack strategies, including an analysis of threat models (\ref{sec:threat_model}) and detailed taxonomies of attacks (\ref{sec:attack_taxo}), such as cybersecurity threats (\ref{sec:syb_sec_attack}) and sensor spoofing attacks (\ref{sec:sensor_spoof_attack}). Section~\ref{adv_attack_llms} focuses on adversarial attacks targeting large language models (LLMs) and large vision-language models (LVLMs), highlighting techniques such as logits-based attacks (\ref{sec:logits_based_attack}), adversarial prompt generation (\ref{sec:adv_prompt_gen}), and cross-modality attacks (\ref{sec:cros——model_attack}). Section~\ref{sec:fail_ai_core} explores challenges and failure modes in embodied AI, including common failure patterns (\ref{sec:cfm}) and algorithm-specific vulnerabilities (\ref{sec:algorithm_specif_fail}). Section~\ref{sec:dataset_taxono_llm_vlm} provides a taxonomy of datasets for evaluating LLMs and LVLMs, encompassing general datasets (\ref{sec:general_dataset}), adversarial datasets (\ref{sec:adv_dataset}), and alignment datasets (\ref{sec:align_dataset}). Finally, Section~\ref{sec:conclud} concludes the article with a synthesis of key findings and proposed directions for future research.

\section{Vulnerability Category}\label{sec:robot_vul}

\begin{figure*}
    \centering
    \includegraphics[width=0.9\linewidth]{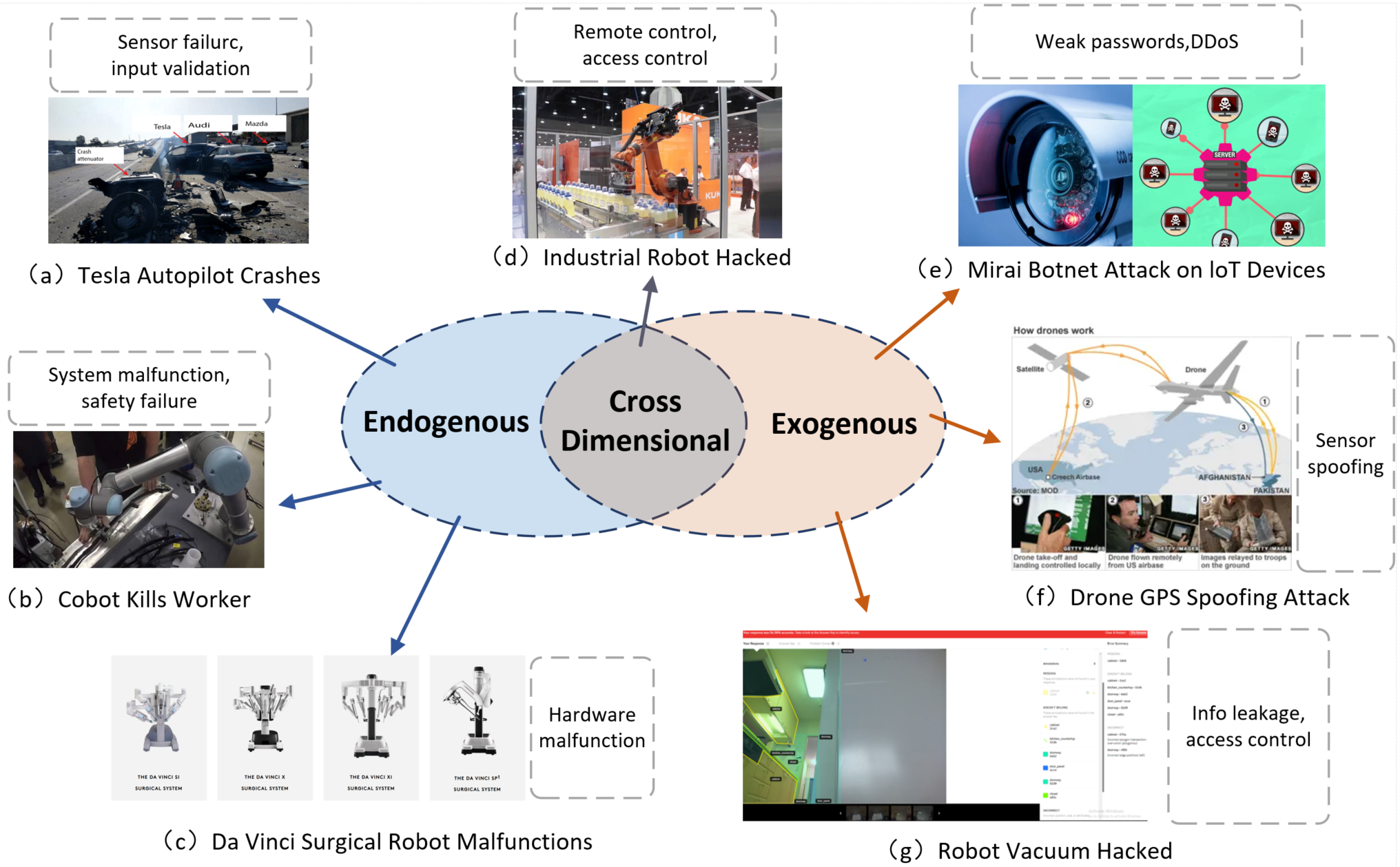}
    \caption{Real-World Risk Cases in Embodied AI.    
    }
    \label{fig:real_world_risk}
\end{figure*}

The risks inherent to embodied AI systems can be broadly classified into three key dimensions: Exogenous Vulnerability (Section \ref{sec:ex_vul}), which originates from external factors such as dynamic environments or adversarial attacks, and Endogenous Vulnerability (Section \ref{sec:end_vul}), which arise from internal system failures, including hardware malfunctions and software vulnerabilities. Additionally, there exists a category of Inter-Dimensional vulnerability (Section \ref{sec:cros_d_vul}), where interactions between external and internal factors exacerbate system fragility.

\subsection{Exogenous Vulnerability}\label{sec:ex_vul}

Exogenous risks arise from the system's interaction with its environment or external malicious actors. In this section, we categorize these risks into Dynamic Environmental Factors (Section \ref{sec:dyn_env_fac}), Physical Attacks (Section \ref{sec:cyb_attack}), Adversarial Attacks (Section \ref{sec:adv_attack}), Cybersecurity Threats (Section \ref{sec:cybersec_threat}), and Human Interaction and Safety Protocol Failures (Section \ref{sec:human_interc_sef_proto}).

\subsubsection{Dynamic Environmental Factors}\label{sec:dyn_env_fac}

Embodied AI systems, especially in dynamic environments, rely on sensor data for perception and interaction \cite{huang2018survey}. However, environmental changes or adversarial perturbations can mislead deep neural networks, leading to misclassifications with safety risks \cite{yeong2021sensor}. Sensor failures in autonomous vehicles, for instance, have been linked to fatal accidents due to misinterpretation of road conditions \cite{tesla_autopilot_crashes}. Moreover, the lack of robust verification mechanisms for deep neural networks in safety-critical applications complicates regulatory compliance and deployment reliability \cite{shahzad2024robust,huang2017safety}. Addressing these challenges requires advancements in sensor fusion, adversarially robust AI models, and rigorous safety verification frameworks to enhance the reliability of embodied AI systems.

\subsubsection{Physical Attacks}\label{sec:cyb_attack}

This attack strategy involves direct hardware tampering, enabling adversaries to manipulate components, disrupt performance, or cause physical damage \cite{9763485}. As cyber-physical systems (CPSs) integrate diverse technologies, ensuring robust security becomes increasingly complex \cite{raval2018competitive}. Furthermore, the rise of Industry 4.0 expands the attack surface, exposing CPSs to novel threats with potential economic and physical consequences \cite{longari2024janus}.

\subsubsection{Adversarial Attacks}\label{sec:adv_attack}

Adversarial attacks present a significant challenge to embodied AI systems, such as autonomous vehicles and robots, by introducing imperceptible perturbations to input data, thereby deceiving deep learning models responsible for real-time perception and decision-making. These attacks, which can be executed in both white-box and black-box settings, are particularly concerning in safety-critical applications like autonomous driving and robotic surgery, where erroneous decisions can result in severe consequences, including physical harm and system failures \cite{szegedy2013intriguing,goodfellow2014explaining,tesla_autopilot_crashes}. Moreover, the physical embodiment of these systems makes them susceptible to adversarial examples crafted for real-world environments, further amplifying the risks \cite{eykholt2018robust}. Although defense mechanisms such as adversarial training have shown promise, ensuring the robustness of embodied systems against such attacks remains an open problem \cite{madry2017towards}. For further details of adversarial attacks, we refer readers to Section \ref{sec:adv_attack_to_vul}.

\subsubsection{Cybersecurity Threats}\label{sec:cybersec_threat}

As embodied AI systems increasingly integrate with the Internet of Things (IoT) and cloud-based infrastructures, they become vulnerable to a wide range of cybersecurity attacks. These attacks can exploit weak security configurations, such as default passwords or unencrypted communications, to gain unauthorized access or disrupt system operations. A notable example is the Mirai botnet attack, which exploited weak passwords in IoT devices to launch a large-scale Distributed Denial of Service (DDoS) attack \cite{mirai_botnet_attack}.  Similarly, drones can be hijacked through GPS spoofing, leading to unauthorized or dangerous actions \cite{drone_gps_spoofing}.

\subsubsection{Human Interaction and Safety Protocol Failures}\label{sec:human_interc_sef_proto}
Collaborative robots (cobots) and drones, designed to operate alongside humans, pose significant safety risks if their safety protocols are compromised. These risks are often exacerbated by human error or malicious intent. For instance, a malfunctioning cobot at a Volkswagen plant led to the death of a worker due to a failure in the robot's safety protocols \cite{cobot_kills_worker}.

\subsection{Endogenous Vulnerability}\label{sec:end_vul}

Endogenous risks originate within the system itself, including hardware failures, software bugs, and design flaws. These risks are often more predictable but can still lead to severe consequences if not adequately mitigated.

\subsubsection{Sensor and Input Validation Failures} \label{sec:sensor_valid_fail}Embodied AI systems rely on a variety of sensors to gather data about their environment. However, sensor malfunctions or improper input validation can lead to incorrect environmental assessments, resulting in unsafe actions. For example, in autonomous vehicles, sensor failures or misinterpretations of sensor data have been linked to accidents, as the system may fail to detect obstacles or misjudge distances \cite{tesla_autopilot_crashes}.

\subsubsection{Hardware and Mechanical Failures}\label{sec:hard_mecha_failure} Mechanical components in embodied AI systems, particularly in industrial and medical robots, are prone to wear and failure over time. These failures can disrupt system operations and, in critical applications, lead to severe consequences. For instance, mechanical malfunctions in surgical robots, such as the Da Vinci Surgical System, have resulted in emergency interventions and, in some cases, the need to convert to open surgery \cite{kim2009failure}. In industrial settings, hardware failures can lead to production delays or safety hazards.

\subsubsection{Software Vulnerabilities and Design Flaws}\label{sec:softwa_vul}

Embodied AI systems are also susceptible to software bugs and design flaws that can compromise their safety and reliability. These vulnerabilities may arise from improper handling of edge cases, insufficient testing in diverse environments, or flaws in decision-making algorithms. For example, a poorly designed control algorithm in an autonomous vehicle could lead to unsafe driving behaviors in complex traffic scenarios. Similarly, software vulnerabilities in industrial robots could be exploited to cause operational disruptions or safety incidents \cite{industrial_robot_hacked}.

\subsection{Inter-Dimensional Vulnerability}\label{sec:cros_d_vul}

Some risks may span both exogenous and endogenous dimensions, where external factors exacerbate internal vulnerabilities. For example, a cyber attack (exogenous) that exploits a software vulnerability (endogenous) could lead to unauthorized control of a robot or vehicle, resulting in physical damage or operational disruption. Similarly, environmental factors such as extreme temperatures or humidity could accelerate hardware degradation, leading to mechanical failures. The remainder of this section will discuss Instruction Misinterpretation (Section \ref{sec:instruc_misint}), Ethical and Safety Implications of Interactive Agents (Section \ref{sec:ethical_interc_ag}), and the Lack of Robustness in Unseen Environments (Section \ref{sec:lack_robost}).

\subsubsection{Instruction Misinterpretation} \label{sec:instruc_misint} One of the primary safety concerns in interactive tasks is the potential for misinterpretation of instructions. For example, in Talk2car \cite{deruyttere2019talk2car} tasks, where agents control self-driving cars based on natural language instructions, a misinterpretation could lead to dangerous driving behaviors or accidents. Similarly, in indoor navigation tasks like ALFRED \cite{shridhar2020alfred}, incorrect object manipulation or navigation could result in damage to the environment or harm to humans if deployed in real-world robotic systems \cite{shridhar2020alfred}.

\subsubsection{Ethical and Safety Implications of Interactive Agents} \label{sec:ethical_interc_ag} In tasks like EQA (Embodied Question Answering), where agents engage in dialogue with users, there is a risk of the agent providing incorrect or misleading information. This could have serious consequences in critical applications such as healthcare or autonomous driving \cite{das2018embodied}. Ensuring that agents are transparent about their limitations and capable of handling uncertainty is crucial for maintaining safety and trust in these systems.

\subsubsection{Lack of Robustness in Unseen Environments} \label{sec:lack_robost} Many VLN tasks, such as Behavioral Robot Navigation, test agents in both seen and unseen environments \cite{anderson2018vision}. However, agents often struggle to generalize to new, unseen environments, which could lead to failures in navigation or task completion. In real-world applications, this could result in agents becoming stuck, causing delays, or even creating hazardous situations if the agent is unable to adapt to unexpected obstacles or changes in the environment.

\section{Attack to Vulnerabilities}\label{sec:adv_attack_to_vul}

The advent of multimodal large models, particularly LVLMs derived from LLMs, has revolutionized embodied AI by enabling advanced capabilities in perception, reasoning, and interaction across vision and language tasks. These models leverage extensive pretraining, high capacity, and sophisticated alignment techniques to achieve state-of-the-art performance. However, their multimodal nature—integrating both visual and textual modalities—significantly expands the attack surface, introducing unique vulnerabilities. Adversaries can exploit visual inputs, textual prompts, or their interactions to craft sophisticated attack vectors, exposing critical gaps in existing defenses.

This section systematically examines the vulnerabilities of these systems, beginning with a Vulnerability Analysis (Section \ref{sec:reason_beh_LLM_VLM}) that highlights the expanded attack surface, adversarial vulnerabilities, and challenges arising from transitions between text and actions. Next, the Threat Model (Section \ref{sec:threat_model}) outlines attacker capabilities and potential targets, followed by an Attack Taxonomy (Section \ref{sec:attack_taxo}) that categorizes threats into three overarching types: exogenous, endogenous, and inter-dimensional vulnerability-centric attacks.

Subsequent sections delve into specific attack methodologies, including Cybersecurity Threats (Section \ref{sec:syb_sec_attack}), Sensor Spoofing Attacks (Section \ref{sec:sensor_spoof_attack}), and Adversarial Attacks (Section \ref{adv_attack_llms}).

\subsection{Vulnerability Analysis}\label{sec:reason_beh_LLM_VLM}

The vulnerabilities of LLMs and LVLMs arise from fundamental limitations in their training paradigms, data coverage, and architectural properties. This section examines three key areas of concern: Training and Data Limitations in Section \ref{sec:train_paradigm}, {Adversarial Vulnerabilities} in neural networks in Section \ref{sec:adv_vul_analysis}, and {Expanded Attack Surface} in multimodal systems in Section \ref{sec:expand_attack_surf}.

\subsubsection{Training and Data Limitations}\label{sec:train_paradigm}
LLMs' autoregressive training paradigm focuses on next-word prediction, which diverges from the goal of generating helpful, truthful, and harmless responses. Safety considerations are not inherently embedded, and fine-tuning with safety datasets offers limited mitigation. Pre-training on uncurated internet data \cite{carlini2024alignedneuralnetworksadversarially} introduces biases \cite{dixon2018measuring} and toxic content \cite{welbl2021challenges}, reinforcing stereotypes or generating harmful outputs. These limitations in training paradigms and data exacerbate the model's vulnerability to adversarial manipulation and misuse.

\subsubsection{Adversarial Vulnerabilities}\label{sec:adv_vul_analysis}
Deep neural networks (DNNs), despite their nonlinear architecture, often exhibit near-linear behavior in high-dimensional input spaces, making them susceptible to adversarial examples \cite{carlini2017towards}. Small input perturbations can cause significant prediction shifts, as demonstrated by attacks like FGSM and Carlini \& Wagner (CW) \cite{carlini2017towards}. Additionally, incomplete input space coverage during training \cite{papernot2016transferability} leads to blind spots, increasing overfitting and reducing generalization \cite{nguyen2015deep}. Near decision boundaries, steep gradients further amplify this vulnerability \cite{papernot2016limitations}. DNNs are also highly sensitive to high-frequency components \cite{szegedy2013intriguing}, creating a mismatch between human perception and model behavior that adversaries can exploit.

\subsubsection{Expanded Attack 
Surface}\label{sec:expand_attack_surf}

LVLMs process multimodal inputs (e.g., text and vision), enhancing capabilities but introducing new vulnerabilities. Adversarial signals in one modality can propagate across others, amplifying their impact \cite{liu2024surveyattackslargevisionlanguage}; for example, manipulated visuals can disrupt textual reasoning. The growing adoption of LVLMs, especially in embodied AI and sensor-driven systems, expands the attack surface. Zhang et al. \cite{zhang2024visualadversarialattackvisionlanguage} demonstrated that adversarial attacks on LVLMs pose significant risks in autonomous driving.

\subsubsection{Transitions from Text to Actions}
Traditional jailbreak attacks, which are designed to bypass LLM safety mechanisms, may not be fully applicable to embodied systems. Embodied LLMs must not only generate text but also plan and execute actions in the physical world. This requires a new attack paradigm that takes into account the unique challenges of action planning and execution in embodied systems.
For example, embodied LLMs often generate structured outputs, such as JSON or YAML, which are then used by downstream control modules to execute actions \cite{qin2023toolllm,wang2024executable,goel2023llms}. Attackers could exploit these structured outputs to manipulate the system's behavior, leading to unsafe or unintended actions.

\subsection{Threat Model}\label{sec:threat_model}

The framework of an adversarial threat model is defined by two primary aspects: the attacker's abilities and their intended goals.

\subsubsection{Attacker Capabilities}\label{sec:attack_cap}

\begin{itemize}
    \item In white-box attacks, adversaries have full access to the system's architecture, parameters, and APIs, as is common in open-source embodied AI systems or simulators \cite{brohan2022rt, shridhar2023perceiver}. Examples of classic white-box attacks include FGSM \cite{p55}, PGD \cite{p47}, APGD \cite{Cui_2024_CVPR}, and CW \cite{carlini2017towards}. This enables the creation of highly targeted and sophisticated attacks. 
    \item {Gray-box Attacks}: Gray-box attacks arise when adversaries have partial access, typically via high-level APIs or external interfaces, but lack control over lower-level components \cite{liang2023code}. These attacks often exploit vulnerabilities in external inputs like sensor data or user commands.
    \item {Black-box Attacks}: In black-box attacks, adversaries lack knowledge of the system's internals and interact only through input queries. Despite proprietary protections in commercial systems, attackers can still exploit external inputs to cause harm. 
\end{itemize}

\subsubsection{Attack Targets}\label{sec:attack_target}
The specific targets of an attack can vary depending on the attacker's goals. Common targets in embodied AI systems include:
\begin{itemize}

    \item Perception Systems: Attacks on sensors, such as cameras, LiDAR, and GPS (discussed in Section \ref{sec:sensor_spoof_attack}), can disrupt a system's ability to accurately interpret its environment, resulting in faulty decision-making. For example, sensor spoofing can cause an autonomous vehicle to misjudge distances or fail to detect obstacles. The Drone GPS Spoofing Attack \cite{drone_gps_spoofing} is a notable example, where a civilian drone was hijacked via GPS spoofing, leading to illegal use.

    \item Control Systems: Attacks on the system's control algorithms' vulnerabilities (discussed in Section \ref{sec:fail_ai_core}) can lead to unsafe actions, such as erratic movements or failure to follow safety protocols. For instance, an attacker could manipulate a robot's control system to cause it to collide with objects or humans. 

    \item Communication Channels: Embodied AI systems often rely on wireless communication for coordination and control. Attacks on these communication channels, such as jamming or Man-in-the-Middle Attacks (MitM) (discussed in Section \ref{sec:syb_sec_attack}), can disrupt the system's ability to receive critical updates or commands.

\end{itemize}

\begin{table}[t]
\centering
\caption{Comparison of Input Manipulation Attacks.}
\label{tab:comp_input_manipul_attack}
\scriptsize 
\renewcommand{\arraystretch}{1.2} 
\begin{tabularx}{\linewidth}{p{1.8cm} X X X X X}
\toprule
\textbf{Aspect}        & \textbf{Adversarial Attacks} & \textbf{Sensor Spoofing Attacks} & \textbf{Command Injection} & \textbf{Prompt Injection} & \textbf{Adversarial Prompting} \\ 
\midrule
\textbf{Domain}        & Digital                     & Physical                         & Physical                   & Digital                   & Digital                         \\ 
\textbf{Target}        & Model predictions           & Sensor data pipeline             & Action execution layer     & Response generation layer & Response generation layer       \\ 
\textbf{Mechanism}     & Input perturbations         & Environmental manipulation       & Malicious commands         & Malicious prompts         & Semantic prompt manipulation   \\ 
\textbf{Impact}        & Misclassification           & Physical consequences            & Physical consequences      & Harmful/unintended outputs & Harmful/unintended outputs      \\ 
\textbf{Scope}         & Broad (all ML models)       & Embodied AI systems              & NLP-based systems          & LLMs/LVLMs                & LLMs/LVLMs                      \\ 
\bottomrule
\end{tabularx}
\end{table}

\subsection{Attack Taxonomy}\label{sec:attack_taxo}
Attacks to embodied AI can be systematically categorized based on attacker capabilities, target components and vulnerabilities, and their overall impact on system performance. This taxonomy encompasses a range of strategies, including data poisoning, sensor spoofing, and adversarial perturbations, each with distinct mechanisms and implications. In this section, we provide a comprehensive overview of these attack types:

\begin{itemize}

    \item {Exogenous Vulnerability Centric Attacks}
    \begin{itemize}
        \item {Data-Centric Attacks}
        \begin{itemize}
            \item {Data Poisoning}: Attackers inject malicious data into the training set, causing the model to learn incorrect associations, potentially leading to failures in safety-critical applications like autonomous driving or robotic surgery.
        \end{itemize}
        \item {Input Manipulation Attacks}
        \begin{itemize}
            \item Adversarial Attacks: Adversarial attacks are a significant threat to machine learning models, particularly those used in perception systems. Szegedy et al. \cite{szegedy2014intriguingpropertiesneuralnetworks,ren2019security} first demonstrated that many machine learning algorithms are vulnerable to adversarial examples, which are small perturbations in input data that lead to incorrect model predictions. This vulnerability is largely due to the inherent non-linearity of neural networks \cite{goodfellow2015explainingharnessingadversarialexamples}.

            \item Sensor Spoofing Attacks: These attacks exploit vulnerabilities in the sensor data pipeline of embodied AI systems, targeting the processes of data acquisition, processing, and interpretation. Unlike adversarial attacks, which manipulate digital inputs to deceive machine learning models, sensor spoofing attacks occur in the physical domain, manipulating the environment or sensor signals directly. For example, attackers can inject false data into sensors like LiDAR, GPS, or cameras, causing robots to misinterpret their surroundings and exhibit unintended behaviors, such as collisions or navigation failures \cite{xu2023sok, zhou2023robust}, misleading autonomous vehicles \cite{xie2023reasoning}, or embedding malicious voice commands to exploit smart assistants for unauthorized actions.
            
            \item {Command Injection}: In NLP-based systems like voice assistants or robots, attackers can inject malicious commands via voice or text inputs, potentially triggering unintended actions such as unlocking doors, disabling security systems, or performing dangerous maneuvers.
            
            \item Jailbreak Attacks: Jailbreak attacks on LLMs and LVLMs involve crafting prompts that bypass the model's safety mechanisms, posing significant risks to information security. In contrast to command injection attacks, which primarily target physical systems, jailbreak attacks exploit vulnerabilities in the model's response generation. Two representative techniques include:
            \begin{itemize}
                \item {Prompt Injection}: Manipulating the model's output by injecting malicious prompts that cause the LLM to generate harmful or unintended responses.
                \item {Adversarial Prompting}: Introducing subtle perturbations to the input prompt to elicit unintended responses from the LLM.
            \end{itemize}
        \end{itemize}
        A comparison of different input manipulation attacks is shown in Table \ref{tab:comp_input_manipul_attack}.
        \item {System and Infrastructure Attacks}
        \begin{itemize}
            \item {API Manipulation}: Attackers leverage exposed APIs to send malicious commands, causing systems to behave incorrectly, such as manipulating a robot's navigation to take unsafe paths or collide with obstacles.
            \item {Denial-of-Service (DoS) Attacks}: Attackers overwhelm systems with excessive input queries, causing slowdowns or unresponsiveness, which is especially dangerous for real-time systems like autonomous vehicles or drones.
            \item Man-in-the-Middle (MitM) Attacks: Exploiting vulnerabilities in wireless communication, MitM attacks intercept and manipulate data between systems and external services, threatening the safety of autonomous vehicles and robotic systems by tampering with sensor data or control commands \cite{conti2016survey, rouf2010security}. Weak encryption in communication protocols broadens the attack surface, while countermeasures like strong encryption (e.g., TLS, IPsec) and intrusion detection systems (IDS) help mitigate risks \cite{alaba2017internet, singh2011survey}.
        \end{itemize}
    \end{itemize}

    \item {Endogenous Vulnerability Centric Attacks}
    \begin{itemize}
        \item {Model-Centric Attacks}
        \begin{itemize}
        \item Model Extraction Attacks: Attackers replicate a model's functionality or uncover vulnerabilities through methods such as reverse engineering (using full system access) or query-based extraction (approximating the model's decision boundaries through extensive input-output queries).
        \end{itemize}
        \item {System and Infrastructure Attacks}
        \begin{itemize}
            \item {Supply Chain Attacks}: By targeting vulnerabilities in development or deployment, attackers insert malicious code, as seen in the SolarWinds attack, which impacted organizations globally. \cite{wolff2021navigating}
        \end{itemize}
        \item {Sensor and Hardware Attacks}
        \begin{itemize}
            \item {Side-Channel Attacks}: Exploiting indirect data leaks (e.g., timing or power usage), attackers infer sensitive information, posing risks to secure systems like embedded robotics. \cite{kolhe2022lock}
            \item Firmware Attacks: Exploit low-level software in embodied systems (e.g., autonomous vehicles, drones) to gain persistent control, bypassing higher-level security; difficult to detect and mitigate, they demand defenses like secure boot and runtime firmware verification.
        \end{itemize}
        \item {Exploitation of Vulnerabilities}
        \begin{itemize}
            \item {Zero-Day Exploits}: Attackers exploit undisclosed vulnerabilities to execute effective attacks on critical systems, such as healthcare robots or drones.
        \end{itemize}
    \end{itemize}

    \item {Inter-Dimensional Vulnerability Centric Attacks}
    \begin{itemize}
        \item {Sophisticated and Coordinated Attacks}
        \begin{itemize}
            \item {Advanced Persistent Threats (APTs)}: These attackers have the resources and expertise to conduct long-term, stealthy attacks, gradually compromising the system over time without being detected. \cite{alshamrani2019survey}
            \item Ransomware Attacks: These attacks encrypt critical files or lock control software, rendering embodied AI systems (e.g., autonomous vehicles, industrial robots) inoperable until a ransom is paid. Embodied AI systems are particularly vulnerable due to their reliance on real-time data processing, where ransomware-induced failures can cascade across interconnected systems \cite{humayed2017cyber}. 
        \end{itemize}
    \end{itemize}

\end{itemize}

\subsection{Cybersecurity Threat}\label{sec:syb_sec_attack}

Cybersecurity threats have become increasingly sophisticated, targeting the core components of intelligent systems to exploit vulnerabilities and disrupt operations. From intercepting sensitive communications through Man-in-the-Middle Attacks (MitM) (Section \ref{sec:mim_attack}) to Sensor-to-Model Attacks (Section \ref{sec:data_inject_attack}), compromising firmware (Section \ref{sec:firmware_attack}), and exploiting Side-Channel Attacks (Section \ref{sec:side_attack}), these threats undermine system integrity and confidentiality. Additionally, Ransomware Attacks (Section \ref{sec:ransom_attack}) and Supply Chain Attacks (Section \ref{sec:supp_chain_attack}) further amplify risks by targeting critical infrastructure and trusted dependencies. Understanding these attack vectors is essential for developing proactive and resilient embodied AI systems.

\subsubsection{Man-in-the-Middle Attacks (MitM)}\label{sec:mim_attack}

MitM attacks exploit the reliance of embodied systems on real-time communication, particularly over wireless protocols, to intercept and manipulate data exchanged between the system and external services such as cloud servers or remote control units \cite{conti2016survey}. These attacks are especially critical in autonomous systems, where the integrity of real-time data is essential for safe operation. For instance, in autonomous vehicles, MitM attacks can tamper with navigation or sensor data, leading to unsafe rerouting or misinterpretation of the environment \cite{rouf2010security}. Similarly, in robotic systems, such attacks can compromise control commands, resulting in mission failures, production defects, or even physical harm to humans. The growing connectivity of these systems—enabled by inter-agent communication, infrastructure-linked networks in autonomous vehicles, and cloud-based control in robotics—further broadens the attack surface. Weaknesses in communication protocols, particularly those lacking robust encryption, provide adversaries with opportunities to intercept and modify data in transit, enabling malicious outcomes such as data theft, system disruption, or complete system takeover \cite{alaba2017internet}.

To mitigate the risks posed by MitM attacks, several countermeasures are essential. Strong encryption protocols, such as Transport Layer Security (TLS) or IPsec, can help protect data integrity and confidentiality during transmission \cite{singh2011survey}. Moreover, real-time intrusion detection systems (IDS) can monitor network traffic for anomalies that may indicate an ongoing MitM attack, providing an additional layer of defense \cite{da2019internet}.

\subsubsection{Sensor-to-Model Attacks}\label{sec:data_inject_attack}

Sensor-to-model attacks target the integrity of sensor data, which is fundamental to the perception and decision-making processes in embodied AI systems \cite{mo2013detecting}. In these systems, accurate sensor data is crucial for tasks such as navigation, object recognition, and interaction with the physical environment. By injecting falsified or malicious data into the system, attackers can manipulate the system's understanding of its surroundings, leading to potentially dangerous outcomes. This makes sensor-to-model attacks both stealthy and highly effective, especially in real-time systems where rapid decision-making is required \cite{urbina2016limiting}. For example, in autonomous driving, an attacker could inject spoofed LiDAR or camera data, causing the vehicle to perceive obstacles that do not exist, leading to unnecessary evasive maneuvers or even collisions \cite{petit2015remote}. Similarly, in industrial robotics, manipulated sensor data could cause a robot to misinterpret its environment, resulting in production errors or safety hazards.

To mitigate the risks posed by sensor-to-model attacks, robust sensor fusion algorithms and anomaly detection techniques are essential. Sensor fusion, which combines data from multiple sources (e.g., LiDAR, radar, and cameras), can help identify inconsistencies between sensor readings, making it harder for attackers to inject false data into all sensors simultaneously. Additionally, anomaly detection systems can monitor sensor data streams for unusual patterns or behaviors, providing an additional layer of defense against data manipulation. These techniques, when combined with cryptographic methods for securing sensor communication, can significantly enhance the resilience of embodied AI systems against sensor-to-model attacks \cite{sandberg2015cyberphysical}.

\subsubsection{Firmware Attacks}\label{sec:firmware_attack}

Firmware attacks represent a particularly insidious threat to embodied systems, as they target the low-level software that controls hardware components. In embodied AI systems, such as autonomous vehicles, drones, and robots, the firmware governs critical functionalities, including sensor data processing, actuator control, and communication protocols. By compromising the firmware of key components such as sensors or actuators, attackers can gain persistent control over the system, potentially bypassing higher-level security mechanisms. For example, in the case of autonomous drones, a malicious firmware update could alter flight control parameters, leading to erratic or unsafe behavior \cite{costin2014large}. This type of attack could result in significant safety risks, particularly in security-critical applications like autonomous driving or industrial robotics, where precise control and reliability are paramount.

Firmware attacks are notoriously difficult to detect due to their low-level nature and the fact that they often operate beneath traditional security monitoring systems. Once the firmware is compromised, attackers can maintain long-term control over the system, even after reboots or software updates, making detection and remediation particularly challenging. This persistence poses a long-term security risk, necessitating the development of robust defense mechanisms such as secure boot processes, which ensure that only authenticated firmware is loaded during system startup, and runtime firmware verification techniques, which continuously monitor the integrity of the firmware during operation \cite{petit2015remote}.

\subsubsection{Side-Channel Attacks}\label{sec:side_attack}

Side-channel attacks exploit the physical characteristics of a system—such as power consumption, electromagnetic emissions, or timing information—to extract sensitive data \cite{kocher1999differential}. In the context of embodied AI systems, these attacks pose a significant threat, particularly because such systems often rely on cryptographic operations to secure communication and protect sensitive data. For instance, an attacker could monitor the power consumption patterns of a robot during encrypted communication and use that information to infer encryption keys, effectively compromising the system's confidentiality \cite{gandolfi2001electromagnetic}. The susceptibility of embodied systems to side-channel attacks underscores the need for implementing cryptographic algorithms that are resistant to such attacks, thereby ensuring the integrity and privacy of critical operations \cite{mangard2008power}.

\subsubsection{Ransomware Attacks}\label{sec:ransom_attack}
Ransomware attacks pose a severe threat to embodied AI systems, particularly in industrial and operational environments. These attacks typically encrypt critical system files or lock control software, rendering the system unusable until a ransom is paid. In systems like autonomous vehicles or industrial robots, such disruptions can halt entire workflows, leading to significant operational and financial losses. For example, encrypting the navigation system of an autonomous vehicle would immobilize it, causing downtime and potentially putting passengers at risk.

Ransomware attacks have escalated in recent years, with high-profile incidents targeting critical infrastructure and industrial control systems (ICS), such as the 2021 Colonial Pipeline attack \cite{cnn_colonial_2021}. Embodied AI systems, which often rely on real-time data processing and control, are particularly vulnerable because ransomware can disable key functionalities, leading to cascading failures across interconnected systems \cite{humayed2017cyber}. In the case of autonomous systems, the inability to access vital control functions can result in safety-critical failures, especially in high-risk environments like transportation or manufacturing.

To mitigate the impact of ransomware, it is essential to implement robust backup and recovery systems, ensuring that critical data can be restored without paying a ransom \cite{mcbride2018data}. Additionally, proactive detection mechanisms, such as anomaly detection using machine learning, can help identify ransomware activity before it fully compromises a system \cite{scarfone2007guide}.

\subsubsection{Supply Chain Attacks}\label{sec:supp_chain_attack}
Supply chain attacks leverage the complexity of modern hardware and software supply chains, introducing malicious components or code during manufacturing or development. Embodied AI systems, such as autonomous vehicles and robotics, are particularly vulnerable due to their reliance on third-party hardware and software. For instance, an attacker could implant a hardware backdoor in-vehicle sensors, enabling remote control or surveillance of operations. Such attacks are notoriously difficult to detect, as compromised components may remain dormant for extended periods.

The infamous SolarWinds attack exemplifies the potential scale of supply chain compromises, where attackers infiltrated software updates to gain access to numerous organizations globally \cite{wolff2021navigating}. In the context of embodied systems, hardware Trojans—malicious modifications introduced during the manufacturing of integrated circuits (ICs)—pose a significant threat, potentially allowing adversaries to manipulate sensor data or control systems \cite{tehranipoor2010survey}. Furthermore, embedded firmware vulnerabilities in IoT devices, which are increasingly integrated into robotics and autonomous systems, have been shown to be a common vector for supply chain attacks \cite{costin2014large}.

Mitigating these risks requires a multi-layered approach. Hardware attestation techniques can verify the integrity of critical components \cite{guin2014counterfeit}, while secure development lifecycle (SDL) practices ensure that both hardware and software are vetted throughout the supply chain \cite{boyens2022cybersecurity}. Additionally, machine learning-based methods for detecting hardware Trojans offer promise in identifying malicious modifications in ICs before deployment \cite{gubbi2023hardware}.

\subsection{Sensor Spoofing Attacks}\label{sec:sensor_spoof_attack}
\begin{figure*}
    \centering
    \includegraphics[width=\linewidth]{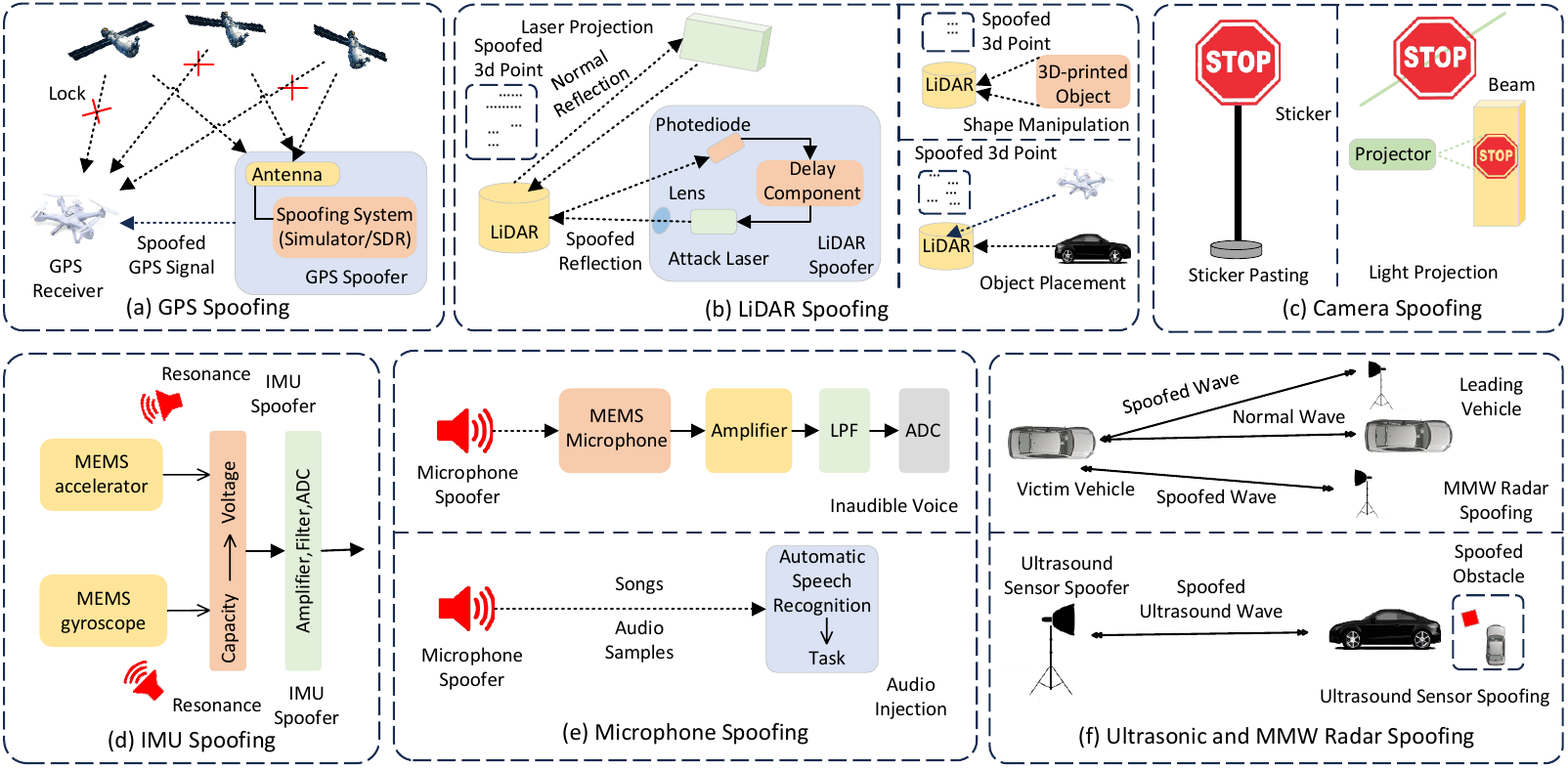}
    \caption{Illustration of Sensor Spoofing Attacks on Six Mainstream Sensors \cite{xu2023sok}.}
    \label{fig:sensor_attack_img}
\end{figure*}

In embodied AI systems, a variety of sensors are employed to enable environmental perception and interaction. These sensors include tactile, vision, proximity, inertial measurement units (IMUs), GPS, and auditory sensors, each contributing unique data to enhance the system's ability to navigate and perform tasks. 
However, despite their critical roles, these sensors are not immune to security vulnerabilities.
Sensor-based attacks on embodied agents have emerged as a critical security concern, as attackers can exploit specific vulnerabilities within the sensor data pipeline to bypass existing safeguards and compromise system integrity \cite{kim2024systematic} as shown in Fig. \ref{fig:sensor_attack_img}.

These attacks target the fundamental processes of sensor data acquisition, processing, and interpretation, often resulting in severe operational failures. For instance, sensor spoofing attacks manipulate a robot's perception by injecting false data into its sensors, which can lead to unintended behaviors such as collisions with obstacles or failure to navigate to the intended destination \cite{xu2023sok, zhou2023robust}. A comprehensive review of these attack strategies and their implications can be found in \cite{xu2023sok}. In the following section, we will provide a detailed analysis of the attack methods specific to each type of sensor.

\subsubsection{Attack on Tactile Sensors}  \label{sec:tactile_sensor_attack}

Tactile sensors are designed to measure force, torque, or pressure distribution across the sensor surface, enabling robots to perceive and respond to physical interactions. They are essential for tasks requiring precise manipulation, such as grasping fragile objects, handling tools, or performing assembly tasks \cite{dahiya2009tactile}. Common types include resistive, capacitive, piezoelectric, and optical sensors, each offering unique advantages in terms of sensitivity, durability, and response time. These sensors are often integrated into robotic systems to provide feedback on grip strength, surface texture, or contact dynamics, allowing robots to adapt their actions in real-time. Recent advancements focus on improving robustness, miniaturization, and reducing wiring complexity by embedding sensors onto flexible PCBs or integrating processing units directly into the sensor array \cite{jiang2021roboticperceptionobjectproperties}. Furthermore, the development of soft and stretchable tactile sensors has expanded their application to soft robotics and wearable devices. Combining tactile sensing with other modalities, such as temperature or vibration, is becoming increasingly important for enhancing the functionality and versatility of robotic systems \cite{mittendorfer2011humanoid}.

Tactile sensors, particularly resistive ones, are prone to physical wear, impacting durability and accuracy. Beyond degradation, they face adversarial attacks that manipulate sensor readings, leading to erroneous robotic responses. Tampering with embedded sensors can disrupt perception, causing misinterpretations \cite{neupane2024security}, while attacks on force or pressure readings further compromise decision-making \cite{tu2022adversarialcontrolloopssensor}. Additionally, cross-talk in capacitive sensors introduces vulnerabilities, enabling signal interference \cite{shi2023embedment}.

\subsubsection{Attack on Vision Sensors}  \label{sec:vision_attack}
Vision sensors, including RGB, RGB-D, and stereo cameras \cite{burrus2011object}, are pivotal in enabling robots to perceive and interact with their environment. These sensors provide rich visual data that, when processed through advanced algorithms like deep learning, SLAM (Simultaneous Localization and Mapping) \cite{zhang2025real}, and optical flow, allow robots to perform complex tasks such as object recognition, navigation, and manipulation. Vision-based systems are widely used in applications like autonomous grasping, real-time mapping, and human-robot interaction, where precise hand-eye coordination and spatial awareness are critical \cite{robinson2023robotic}. Recent developments \cite{ryu2021object} in vision sensors include higher resolution, faster frame rates, and improved depth perception, enabling more accurate and reliable performance in dynamic environments. Additionally, the integration of vision sensors with other modalities, such as tactile or proximity sensors, enhances the robot's ability to understand and interact with its surroundings, even in challenging conditions like low-light or cluttered environments \cite{roberge2023stereotacnovelvisuotactilesensor}.

 However, these vision sensors, which convert optical signals into digital inputs, are particularly susceptible to adversarial manipulations \cite{chakraborty2021survey}. Even slight perturbations in the visual input can result in deep-learning models misclassifying objects or misinterpreting entire scenes. This vulnerability is especially concerning in safety-critical applications, such as self-driving cars or medical robotics, where perception errors can lead to disastrous outcomes \cite{ren2020adversarial}. Adversaries can exploit these weaknesses by introducing subtle visual perturbations that manipulate the sensor's output, leading to incorrect system behavior. Key adversarial techniques include:

\begin{itemize}
    \item {Adversarial Patch (Sticker-Pasting)}: By placing visually abnormal but inexpensive patches on objects, attackers can deceive vision systems into making incorrect classifications. These patches are easy to produce and highly effective in fooling perception models \cite{brown2017adversarial,liu2022segment,liu2018dpatch}.
    
    \item {Light Projection Attacks}: Using devices like laser pointers or projectors, adversarial images can be projected onto real-world surfaces or objects. These attacks are cost-effective, with laser pointers being relatively inexpensive compared to projectors, but both can be mounted on mobile platforms like drones \cite{spoof-camera-light-ccs20,spoof-camera-light-iacr20}.
    
    \item {Exploiting Light Spectrum}: Spoofing attacks can also leverage parts of the electromagnetic spectrum invisible to humans, such as infrared light, to manipulate camera sensors without being noticed by human observers \cite{spoof-camera-light-ccs21,yufeng2023light}.
\end{itemize}

These attacks primarily target the perception systems of embodied agents, leading to significant disruptions in tasks such as object recognition and lane detection. The consequences of these disruptions are particularly severe in autonomous systems where real-time decision-making is critical for safety and operational success \cite{spoof-camera-stick-cvpr18,spoof-camera-stick-woot18,spoof-camera-stick-ccs19,kong2020physgan,wang2021dual,spoof-camera-stick-iclr20,spoof-camera-stick-us21,spoof-camera-stick-us21-cq,spoof-camera-raid,yan2022rolling,spoof-camera-light-ccs20,spoof-camera-light-iacr20,spoof-camera-light-cvpr21,spoof-camera-light-ccs21,spoof-camera-light-us21}. Specific attack vectors include:

\begin{itemize}
    \item {Object Detection Manipulation}: Adversaries can manipulate the perception system to misclassify critical objects, such as interpreting stop signs as speed limit signs, or even fabricating or erasing obstacles. This can lead to hazardous control decisions by autonomous systems \cite{spoof-camera-raid,yan2022rolling,spoof-camera-stick-ccs19,cheng2022physical}.
    
    \item {Lane Detection Tampering}: By altering or projecting false lane markings, attackers can influence autonomous vehicles to deviate from their intended paths, particularly in high-speed environments like highways \cite{spoof-camera-stick-us21,spoof-camera-light-iacr20}.
    
    \item {Object Tracking Disruption}: Continuous adversarial interference can disrupt multi-object tracking systems, leading to incorrect predictions of object trajectories. This can severely affect the system's ability to navigate complex environments in real time \cite{spoof-camera-stick-iclr20}.
    
    \item {Velocity Manipulation}: In drones, spoofing attacks that induce false readings of lateral drift velocity can destabilize flight control systems, increasing the risk of crashes or loss of control \cite{spoof-camera-light-woot16}.
\end{itemize}

\subsubsection{Attack on Proximity Sensors} \label{sec:proximity_sensor_attack}
Proximity sensors, such as ultrasonic, infrared (IR), and LiDAR, are crucial for detecting obstacles, measuring distances, and ensuring safe navigation in robotic systems \cite{li2020lidar}. These sensors emit signals—laser pulses, sound waves, or electromagnetic waves—and calculate the time it takes for the signals to return after reflecting off objects. LiDAR is commonly used in autonomous vehicles and mobile robots for 3D mapping and environmental modeling, providing high-resolution spatial data \cite{li2020deep}. Ultrasonic sensors, on the other hand, are often employed in robotic arms and grippers for precise object detection and proximity sensing, especially in close-range applications \cite{bostelman1989electronics}. 
Infrared sensors are lightweight, cost-effective, and widely used for obstacle detection. They emit infrared radiation and detect reflections, making them effective in low-light or foggy conditions where vision sensors may fail \cite{straub2020detecting}.
Proximity sensors are particularly useful in environments where vision sensors may struggle, such as low-light, foggy, or occluded areas. Recent advancements include the development of compact, low-power sensors with improved range and accuracy, as well as the integration of multiple proximity sensing technologies to enhance reliability and versatility in complex environments \cite{kim2020vision}.

The distinct operational principles of proximity sensors introduce security vulnerabilities, allowing adversaries to inject false data or manipulate readings, jeopardizing robotic safety. Attack vectors like signal interference, spoofing, and adversarial perturbations disrupt sensor functionality and mislead perception. Key exploitation techniques include:
\begin{itemize}
    \item {Signal Interference and Manipulation}:  
    All three types of sensors are prone to signal interference, where attackers can manipulate transmitted or received signals to create false objects or distort the perceived position of real obstacles \cite{spoof-lidar-bhe15, spoof-ultrosonic-iotj18, spoof-radar-tifs21}. For instance, attackers may inject fake signals or alter the timing of signal returns, causing systems to either detect nonexistent obstacles or fail to recognize actual hazards \cite{spoof-lidar-us20, spoof-radar-tifs21}.
    
    \item {Environmental Factors}:  
    Sensor performance can also degrade under environmental conditions such as multipath reflections, variations in surface reflectivity, or adverse weather (e.g., rain, fog, or snow). Attackers can exploit these vulnerabilities to reduce sensor accuracy or create blind spots.
\end{itemize}

Spoofing attacks on proximity sensors pose significant risks, especially for autonomous vehicles and robotics. By introducing false objects \cite{spoof-lidar-camera-sp21, spoof-lidar-ccs21} or altering the perceived positions of real ones, attackers can cause systems to perform abrupt stops \cite{spoof-lidar-us20,spoof-radar-tifs21}, change lanes unexpectedly \cite{spoof-radar-phdthesis}, and fail to detect actual hazards \cite{spoof-lidar-bhe15,spoof-ultrosonic-iotj18}. Sensor-specific attacks include:
\begin{itemize}
    \item {LiDAR}:  
    LiDAR sensors use laser pulses to measure distances. Attackers can exploit their reliance on laser reflections through methods such as:
    \begin{itemize}
        \item \textit{Laser Projection}: By synchronizing with the victim LiDAR using a photodiode, attackers can manipulate the delay of laser pulses to inject fake points into the sensor's point cloud. This results in false objects or distorted object locations, leading to navigation errors \cite{spoof-lidar-bhe15,spoof-lidar-shin,jin2022pla,spoof-lidar-ccs19,spoof-lidar-us20}.
    \end{itemize}
    
    \item {Ultrasonic Sensors}:  
    Ultrasonic sensors, often used in applications like automated parking, emit sound waves to measure distances. They are vulnerable to:
    \begin{itemize}
        \item \textit{Spoofing Attacks}: Manipulating the time-of-flight of sound waves to generate false distance readings, causing the system to detect nonexistent obstacles or miss real ones \cite{spoof-ultrosonic-defcon16,spoof-ultrosonic-iotj18,spoof-ultrosonic-ashes19}.
        \item \textit{Environmental Interference}: Echoes and reflections from nearby surfaces can be exploited to further disrupt obstacle detection.
    \end{itemize}
    
    \item {MMW (Millimeter-Wave) Radar}:  
    MMW radar, which emits electromagnetic waves and is often used alongside LiDAR and cameras for robust obstacle detection, is also susceptible to spoofing. Attackers can manipulate radar signals to distort object detection, even in adverse weather conditions.
\end{itemize}

\subsubsection{IMU Spoofing Attacks}\label{sec:imu_attack}
IMUs provide critical data on orientation, velocity, and acceleration through a combination of accelerometers, gyroscopes, and magnetometers. These sensors are essential for motion tracking, navigation, and control in dynamic environments, enabling robots to maintain balance, stabilize movements, and execute precise maneuvers \cite{ahmad2013reviews}. IMUs are widely used in applications such as drone flight control, humanoid robot locomotion, and autonomous vehicle navigation. To improve accuracy and compensate for issues like sensor drift and noise, IMUs are often integrated with other sensors, such as GPS, barometers, or vision systems, in sensor fusion frameworks \cite{shahcheraghi2025acquisition}. Recent advancements in IMU technology focus on miniaturization, reduced power consumption, and enhanced sensitivity, making them suitable for compact and lightweight robotic platforms \cite{burnett2025imu,wu2025wheel}. Additionally, the use of advanced filtering techniques, such as Kalman filters \cite{afrasiabi2025optimising}, further enhances the reliability of IMU data in real-world applications.

However, these systems are vulnerable to various factors, such as sensor drift, noise, and environmental influences like temperature and vibration, which can degrade their performance. Attackers can exploit these vulnerabilities by introducing noise or manipulating calibration, leading to inaccurate sensor readings and compromised system behavior. A particularly dangerous form of attack is IMU spoofing, where attackers use resonant acoustic interference to manipulate the sensor's mass-spring structure. This is often implemented using a function generator, sound amplifier, and tweeter speaker to target the IMU's resonant frequency, typically below 1 MHz \cite{anlysis-imu-imece07,anlysis-imu-isie07,anlysis-imu-tie11,spoof-imu-us15,spoof-imu-eurosp17,spoof-imu-blackhat17,spoof-imu-asiaccs18,spoof-imu-us18}. 
Such spoofing attacks pose significant threats to embodied AI systems: 
\begin{itemize}
    \item Manipulating linear or angular velocity, leading to loss of control, as seen in attacks like output biasing and side-swing \cite{spoof-imu-eurosp17,spoof-imu-blackhat17}; 
    \item Deceiving object detection systems, causing objects to be misclassified, undetected, or falsely created, which can severely impair the system's perception and interaction with its environment \cite{spoof-imu-sp21}. 
\end{itemize}

\subsubsection{GPS Spoofing Attacks}  \label{sec:gps_attack}
The GPS is widely used for position, velocity, and time estimation, making it essential for navigation and tracking in outdoor environments. GPS enables robots and autonomous systems to determine their location with high precision, facilitating tasks such as route planning, geofencing, and long-distance navigation. 

However, GPS signals can be obstructed \cite{spoof-gps-drone-vlsid18,spoof-gps-car-us18} in urban areas, dense forests, or indoor environments, and the system is vulnerable to jamming and spoofing attacks, the lack of robust signal authentication exacerbates these vulnerabilities \cite{gps-spoof-report}. Spoofing attacks can obscure critical objects in LiDAR detection \cite{spoof-gps-car-iccv21}, misalign traffic light ROIs, causing vehicles to run red lights \cite{spoof-gps-car-autosec21}, or induce dangerous deviations like veering off-road or into oncoming traffic \cite{spoof-gps-car-us20}. 
Mechanism of GPS Spoofing include:
\begin{itemize}
    \item {Signal Disruption:} 
    Attackers use a GPS spoofer to broadcast counterfeit signals with higher power than legitimate satellites, disrupting the victim's connection to authentic signals.
    \item {Manipulation of Perceived Location:}
    Once the victim locks onto the spoofed signals, attackers can manipulate the perceived location by altering the pseudorange values and modifying navigation messages \cite{spoof-gps-sensor-ion08,spoof-gps-sensor-ccs12}.
\end{itemize}
Types of GPS Spoofers include:
    \begin{itemize}
        \item {GPS Simulators:} These are powerful devices but are often bulky and expensive.
        \item {Software Defined Radios (SDRs):} These are more affordable, flexible, and easier to use. For example, a pen-sized SDR spoofer was successfully used to mislead autonomous vehicles in \cite{spoof-gps-car-us18}, where the spoofer was concealed in an adversarial vehicle, making detection particularly challenging.
    \end{itemize}

To overcome the limitations of standalone GPS, multi-sensor fusion integrates GPS with IMUs, LiDAR, and vision systems, enhancing positioning accuracy and reliability. GPS-IMU fusion combines global positioning with relative motion data for robust navigation \cite{alaba2024gps}. Advances include high-precision RTK GPS, achieving centimeter-level accuracy via carrier-based ranging and real-time corrections \cite{ekaso2020accuracy}, and GPS-SBAS integration for improved performance in challenging environments \cite{gupta2025satellite}.

\subsubsection{Auditory Spoofing Attacks}\label{sec:audit_attack}

Auditory sensors, especially MEMS microphones, are integral to voice-controlled interfaces in embodied AI systems, such as autonomous vehicles and service robots \cite{spoof-voice-wc19}. These systems rely on auditory input to interpret and execute user commands, enabling seamless human-robot interaction for tasks like navigation or object manipulation. The MEMS microphone architecture, consisting of a transducer, amplifier, low-pass filter (LPF), and analog-to-digital converter (ADC), processes sound by converting air pressure changes into electrical signals, which are then amplified, filtered, and digitized \cite{voice-secon16}.

However, in the context of embodied AI, these auditory sensors introduce significant security and reliability concerns. Ultrasonic attacks, for instance, can exploit signals beyond the human hearing range to inject malicious commands, potentially causing dangerous or unintended behaviors in autonomous systems. Furthermore, environmental noise in dynamic, real-world settings can degrade sensor accuracy, leading to misinterpretation of commands and system malfunctions. Mechanism of auditory spoofing include:
\begin{itemize}
    \item {Inaudible Voice Attacks:} These attacks convert high-frequency ultrasonic signals into inaudible ranges through non-linear demodulation \cite{spoof-voice-inaudible-ccs17}. Such signals can inject malicious commands that are imperceptible to humans but interpretable by auditory sensors. For example, Yan et al. \cite{spoof-voice-inaudible-tdsc21} demonstrated how inaudible voice attacks could manipulate various in-car features of an Audi autonomous vehicle.

    \item {Audio Injection Attacks:} Malicious commands are embedded within seemingly normal audio signals, making them difficult to distinguish from legitimate inputs \cite{spoof-voice-inaudible-tdsc21}. These attacks can be executed using tools ranging from basic media players to sophisticated function generators. For example: Zhou et al. \cite{spoof-voice-wc19} showed that audio injection could be used to control navigation functions in autonomous driving systems.
\end{itemize}

\subsection{Adversarial Attack to LLMs and LVLMs}\label{adv_attack_llms}

LLMs and LVLMs have demonstrated remarkable capabilities in natural language processing and multimodal tasks. However, their susceptibility to jailbreak attacks—where adversaries manipulate the models to generate harmful or unethical outputs—remains a critical concern. This section systematically categorizes these attacks into white-box and black-box scenarios. White-box attacks include logits-based methods (Section \ref{sec:logits_based_attack}) and fine-tuning-based approaches (Section \ref{sec:fintuning_based_attacks}), while black-box attacks focus on Adversarial Prompt Generation (Section \ref{sec:adv_prompt_gen}), encompassing techniques such as Pre/Suffix-based Attacks, Prompt Rewriting, Template Completion, LLM-generated Attacks, Text Embedding Refinement, and Retrieval-Augmented Generation (RAG)-based Jailbreaks. Additionally, we explore Cross-Modality Attacks (Section \ref{sec:cros——model_attack}), Attack Transferability (Section \ref{sec:attack_transfer}), Evaluation Strategies (Section \ref{sec:evaluation_strage}), and Safety Mitigation Techniques (Section \ref{sec:safetyu_mitigation}).

\subsubsection{Logits-based Attacks}\label{sec:logits_based_attack}
Logits-based attacks primarily target the decoding process, influencing token selection to produce harmful or misleading content. These methods iteratively adjust the logits to align the token distribution with the attacker's objectives, effectively steering the model's output. While highly effective, these methods often risk compromising the naturalness and coherence of the generated text. Recent methods include:
\begin{itemize}
    \item {COLD-Attack: Energy-based Constrained Decoding:} Guo et al.~\cite{guo2024cold} introduced COLD-Attack, a controlled text generation framework that automates and optimizes jailbreak prompt creation. COLD-Attack demonstrates high success rates against models such as ChatGPT, Llama-2, and Mistral.
    \item Logit-Based Watermarking: Wong et al. \cite{wong2025an} introduced a logits perturbation method for watermarking LLM-generated text, enhancing detection robustness and text quality. However, it remains vulnerable to attacks that manipulate logits to remove or alter watermarks, risking unauthorized content use.
    \item VT-Attack: Wang et al. \cite{wang2024breakvisualperceptionadversarial} proposed VT-Attack, which targets encoded visual tokens in LVLMs to create adversarial examples. This leads to misinterpretations in visual perception, causing incorrect or harmful outputs, exposing vulnerabilities in LVLMs' visual components.
\end{itemize}

\subsubsection{Fine-tuning-based Attacks}\label{sec:fintuning_based_attacks}
Fine-tuning-based attacks involve retraining a target model on malicious or carefully crafted datasets, thereby increasing its susceptibility to adversarial inputs.
Recent advancements include:
\begin{itemize}
\item {Minimal Fine-tuning:} Qi et al.\cite{qi2023fine} demonstrated that fine-tuning LLMs with a small number of harmful examples can significantly undermine their security. Notably, even predominantly benign datasets may inadvertently degrade model safety during the fine-tuning process.
\item {Oracle-based Fine-tuning:} Yang et al.\cite{yang2023shadow} proposed constructing fine-tuning datasets by querying an oracle LLM with malicious prompts. Models fine-tuned on these datasets exhibited heightened vulnerability to jailbreak attempts.
\end{itemize}

\subsubsection{Adversarial Prompt Generation}\label{sec:adv_prompt_gen}

This section categorizes key techniques of adversarial prompt generation, including pre/suffix-based attacks, prompt rewriting, template completion, LLM-generated attacks, embedding refinement, and RAG-based jailbreaks, highlighting their methodologies and security implications.

\begin{itemize}

\item{Pre/Suffix-based Attacks}: Pre/suffix attacks utilize gradients to generate adversarial prefixes or suffixes, guiding the model toward harmful outputs. These attacks are analogous to adversarial examples in text generation tasks.

\begin{itemize}
    \item  {Greedy Coordinate Gradient (GCG):} 
    Zou et al. \cite{p21} introduced GCG, a gradient-based jailbreak attack that iteratively optimizes a discrete adversarial suffix by computing top-$k$ gradient-based replacements, randomly sampling tokens, and updating with the best replacement. GCG effectively transfers to black-box models like ChatGPT, Bard, and Claude.
    \item {Adversarial Suffix Embedding Translation Framework (ASETF):} Wang et al.~\cite{wang2024asetf} proposed ASETF, which optimizes a continuous adversarial suffix in the embedding space and translates it into a human-readable form via embedding similarity.
    \item {AutoDAN:} Zhu et al.~\cite{zhu2023autodan} proposed AutoDAN, an interpretable gradient-based jailbreak attack. Using the Single Token Optimization (STO) algorithm, AutoDAN sequentially generates semantically meaningful adversarial suffixes capable of evading perplexity-based filters.
\end{itemize}

Gradient-based pre/suffix attacks reveal the intricacies of manipulating model inputs to trigger specific responses but often generate unnatural text. Defenses targeting such high-complexity inputs can mitigate their impact.

    \item{Prompt Rewriting Attacks}: Prompt rewriting attacks involve modifying prompts to exploit the model's vulnerabilities in underrepresented scenarios. These methods include cryptographic, linguistic, and genetic strategies.
    \begin{itemize}
        \item {Cryptographic Strategies:} Yuan et al.~\cite{yuan2024cipherchat} introduced CipherChat, which uses encrypted prompts to bypass content moderation.
    
        \item {Linguistic Strategies:} Deng et al.\cite{deng2023multilingual} and Yong et al.\cite{yong2023low} showed that translating unsafe prompts into low-resource languages effectively bypasses LLM safety mechanisms, exposing cross-lingual vulnerabilities.
    
        \item {Genetic Strategies:} Liu et al.~\cite{liu2024autodan} developed AutoDAN, a hierarchical genetic algorithm for generating stealthy jailbreak prompts.
        
     \item Heuristic Text Search for Sensitive Word Substitution: SurrogatePrompt \cite{ba2024surrogateprompt} exploits vulnerabilities in text-to-image models (e.g., Midjourney) by leveraging the cognitive gap between human-perception-based filters and models trained on large-scale data. Similarly, Gao et al. \cite{gao2024htsattackheuristictokensearch} uses heuristic search and random mutations to generate adversarial prompts by removing sensitive keywords.
         
    \end{itemize}

    \item {Template Completion Attacks}: Template completion attacks involve designing sophisticated templates that exploit a model's inherent capabilities—such as role-playing, contextual understanding, and code execution—to bypass its safety mechanisms.
    
    \begin{itemize}
    \item {Scenario Nesting:} By embedding malicious intent within seemingly harmless contexts, attackers craft deceptive scenarios to manipulate the model into performing restricted actions.
    Lin et al. \cite{lin2024figure} propose the Analyzing-based Jailbreak (ABJ) method, which takes advantage of LLMs' sophisticated reasoning abilities to uncover weaknesses. Li et al. \cite{li2023deepinception} introduce DeepInception, leveraging LLMs' personification abilities to create virtual, nested scenarios for adaptive usage control evasion. Additionally, Ding et al. \cite{loughran2009wolf} develop ReNeLLM, a jailbreak framework that employs a two-phase strategy—Scenario Nesting and Prompt Rewriting—to generate effective jailbreak prompts.
    
    \item {Contextual Learning:} Adversarial inputs are seamlessly embedded within the context, influencing the model to generate unintended or harmful outputs.
    Wei et al. \cite{wei2023jailbreak} present the In-Context Attack (ICA) approach to alter the behavior of language models. Deng et al. \cite{deng2024pandora} explore indirect jailbreak methods in Retrieval-Augmented Generation (RAG) scenarios, where external knowledge sources are integrated with LLMs like GPTs. Furthermore, Li et al. \cite{li2023multi} propose Multi-step Jailbreak Prompts (MJP) to investigate the extraction of Personally Identifiable Information (PII) from models such as ChatGPT.
    \item {Code Injection:} Malicious code snippets are introduced to exploit the model's programming and execution capabilities, enabling harmful outputs or unintended behaviors. 
    Kang et al. \cite{kang2024exploiting} and Lv et al. \cite{lv2024codechameleon} leverage programming language features to develop tailored jailbreak instructions for targeting LLMs.
    \end{itemize}

    \item{LLM-generated Attacks}: LLM-generated attacks utilize the generative capabilities of LLMs to simulate attackers, enabling the automated and efficient creation of adversarial prompts. These methods leverage the model's own strengths to refine and optimize attack strategies.
    
    \begin{itemize}
    \item {Persuasive Adversarial Prompts (PAP):} Zeng et al.\cite{zeng2024johnny} trained LLMs to generate persuasive adversarial prompts by incorporating a taxonomy of persuasion techniques, enhancing the effectiveness of the attacks.
    \item {Prompt Automatic Iterative Refinement (PAIR):} Chao et al.\cite{chao2023jailbreaking} proposed \textit{PAIR}, a collaborative framework where multiple LLMs iteratively refine jailbreak prompts, achieving higher success rates through cooperative optimization. Dong et al. \cite{dong2024jailbreaking} propose a multi-agent system where a mutation agent generates adversarial prompts and a selection agent evaluates and refines them. Using contextual learning and chain-of-thought reasoning, the system iteratively improves by learning from successes and failures.
    \end{itemize}

    \item{Text Embedding Refinement Techniques}: Ma et al. \cite{ma2024jailbreaking} generate adversarial concept embeddings by calculating the difference between antonyms' embeddings. By adding these embeddings to the original prompt embeddings,  generate new embeddings containing NSFW concepts, effectively bypassing the model's safety mechanisms to produce harmful content.
    \item{Retrieval Augmented Generation (RAG) based Jailbreak}: PANDORA \cite{deng2024pandora} introduces an indirect attack on LLMs using RAG. By poisoning external knowledge bases with crafted content, it manipulates the RAG process to achieve jailbreaks with higher success rates than direct methods.
\end{itemize}

\subsubsection{Cross-Modality Attack }\label{sec:cros——model_attack}
Cross-modal jailbreaks transfer attacks across text, image, and audio modalities. Text-to-image methods exploit LVLMs via role-playing, logical flowcharts, low-resource languages, benign term substitution, or embedding jailbreak text in images to bypass filters and generate harmful content. Text-to-audio methods convert textual jailbreaks into adversarial audio prompts, leveraging role-playing or fictional narratives.
\begin{itemize}
    \item {Text-Image Joint Prompt Optimization Strategy}:
This strategy jointly optimizes textual and visual prompts to address single-modality attack limitations. Niu et al. \cite{niu2024jailbreaking} present a jailbreak attack on LVLMs using a maximum likelihood estimation-based algorithm to generate three types of prompts: (1) Image Jailbreak Prompts (imgJP), which directly induce inappropriate responses; (2) Perturbation Jailbreak Prompts (deltaJP), crafted to generalize across unseen images by adding minimal perturbations; and (3) Embedding Jailbreak Prompts (embJP), which reverse embeddings to produce transferable text-based prompts (txtJP).

\item Text-Embedded Adversarial Image Generation: FigStep\cite{gong2023figstep} exploits the capability of LVLMs to comprehend textual instructions embedded within images by encoding harmful content directly into the visual modality. By pairing these adversarially crafted images with benign textual prompts, FigStep effectively manipulates the VLM to generate detailed and potentially harmful outputs.
\item  Fictional Storytelling for Text-to-Voice Jailbreaking: VoiceJailbreak \cite{shen2024voice} introduces an innovative attack method that leverages the principles of fictional storytelling—such as crafting compelling backgrounds, characters, and plots—to "humanize" GPT-4o and persuade it to bypass its safety mechanisms. This approach significantly enhances the success rate of attacks by generating simple, \textbf{audible}, and highly effective jailbreak prompts.
\item  Visual-RolePlay (VRP): VRP \cite{ma2024visual} exploits LVLM vulnerabilities by generating high-risk character descriptions with negative traits via LLMs, converting them into visual representations, and pairing these images with benign role-playing instructions to induce malicious outputs.
\end{itemize}

\subsubsection{Attack Transferability}\label{sec:attack_transfer}
Attack transferability is a critical issue in adversarial research on Large Language Models (LLMs) and Large Vision-Language Models (LVLMs). The following studies explore different aspects of attack transferability across models:
\begin{itemize}
    \item Attack Transferability Across LVLMs: Rylan et al. \cite{schaefferfailures} demonstrate that gradient-based universal image jailbreaks exhibit limited transferability across different LVLMs. However, partial transferability is observed in two cases: (1) between LVLMs trained on the same dataset with similar initialization, and (2) across different training checkpoints of the same VLM. Notably, targeting a broader set of "highly similar" LVLMs significantly enhances transferability to a specific target model. Key factors influencing transferability include shared visual backbones and language components. 
   Chung et al. \cite{chung2024transferableattacksvisionllmsautonomous} introduced a typographic attack framework against Vision-LLMs in autonomous driving. Their method generates misleading answers that disrupt reasoning, showing high attack transferability across models like LLaVA, Qwen-VL, VILA, and Imp, highlighting security risks for autonomous vehicles.
    \item He et al. \cite{he2024tubacrosslingualtransferabilitybackdoor} studied cross-lingual backdoor attacks in multilingual LLMs, showing that poisoning data in one or two languages can impact outputs in others. Their experiments on mT5, BLOOM, and GPT-3.5-turbo achieved over 95\% attack success rates, with larger models being more vulnerable.
    \item Transferability of LoRA-Based Attacks: Liu et al. \cite{liu2024loraasanattackpiercingllmsafety} investigated LoRA-based backdoor attacks in LLMs, showing that backdoors persist even with multiple LoRA modules. They also analyzed attack transferability, revealing security risks in sharing and using these adaptation modules.
\end{itemize}

\subsubsection{Evaluation Strategies}\label{sec:evaluation_strage}
Evaluating Embodied AI with LLMs and LVLMs requires comprehensive strategies, with recent research proposing new frameworks and benchmark datasets:
\begin{itemize}
\item {Red Team Framework for Large Visual Language Models}: Chen et al. \cite{chen2024red} proposed a red team framework combining visual and textual jailbreak prompts to create multimodal attacks. By introducing entropy and novelty rewards, it enhances test case diversity and uncovers VLM vulnerabilities in harmful content generation.
\item{Multimodal Benchmark Dataset for VLM Evaluation}: To evaluate VLM security, Ying et al. \cite{ying2024unveiling} constructed a multimodal benchmark dataset using sources like AdvBench and SafeBench. Data is cleaned, categorized by safety policies (e.g., violence, sexual content), and paired with adversarial prompts generated via LLM attack methods. Text prompts are combined with blank, noise, or natural images to create text-image attack cases, forming a comprehensive multimodal evaluation dataset.
\item
Evaluating Abilities of LVLMs: Saito et al. \cite{hayashi2024irrimagereviewranking} examined LVLMs' ability to generate image review texts beyond captioning, assessing composition and exposure. They proposed a rank correlation-based evaluation method comparing human and LVLM rankings and introduced a benchmark dataset, showing that some models effectively differentiate review quality.

\item{
Zhang et al. \cite{zhang2024talecteachllmevaluate} proposed TALEC, a model-based evaluation method that customizes criteria for specific domains using in-context learning. By integrating zero-shot and few-shot learning, TALEC enhances focus on relevant information, achieving over 80\% correlation with human judgments in some tasks.
}
\end{itemize}
\text

\subsubsection{Safety Mitigation}\label{sec:safetyu_mitigation}
Ensuring the safety of Embodied AI with LLMs and LVLMs is crucial, with recent research proposing various mitigation strategies:
\begin{itemize}
    \item Defensive Training and Model Optimization:
        \begin{itemize}
        \item {Forgetting Unsafe Content in Multimodal Models}: To enhance LVLMs security, research explored methods for forgetting harmful content in text and multimodal domains. By using gradient ascent to increase the loss for harmful sample generation and gradient descent to reduce the loss for benign sample generation, the model can effectively forget harmful content. In cross-modal safety alignment studies, it was found that forgetting harmful content in the text domain alone significantly reduces the attack success rate of LVLMs. However, multimodal forgetting does not provide additional advantages and requires significantly more computational resources. This highlights the importance of text-domain safety optimization in improving VLM security.

        \item In-context Defense (ICD) \cite{wei2023jailbreak} improves model robustness by showcasing examples of rejecting harmful inputs. This method utilizes the in-context learning (ICL) capabilities of LLMs to guide the model in learning appropriate rejection behaviors.
    \end{itemize}

    \item Content Filtering and Detection:
    \begin{itemize}
     \item The OpenAI Moderation Endpoint API \cite{openai2023moderation} is a content moderation system developed by OpenAI. It uses a multi-label classification approach to assign responses to 13 distinct categories. Any response identified within these categories is marked as a violation of OpenAI's policy guidelines \cite{openai2024usage}.
    
    \item The Perplexity Filter (PPL) \cite{jain2023baseline} aims to detect incoherent attack prompts by setting a perplexity threshold. It leverages another language model to evaluate the perplexity of either the entire prompt or its segments, discarding prompts that exceed the threshold.
    \end{itemize}
    
   \item Input Perturbation and Prediction Aggregation:
    \begin{itemize}
    \item SmoothLLM \cite{zhang2023defending} employs a two-step process: first, it generates perturbed versions of input prompts; second, it aggregates the predictions from these variations to produce a consolidated result.
\end{itemize}

\end{itemize}

\begin{table*}[htbp]
\centering

\caption{Summary of Robotics-Specific Datasets with Emphasis on Object Manipulation, Robotic Grasping, and Control Tasks.}
\label{tab:robotics_datasets}
\resizebox{\textwidth}{!}{
\begin{tabular}{|p{3cm}|p{5cm}|p{3cm}|p{1.5cm}|p{3.5cm}|}
\hline
\textbf{Dataset Name} 
& \textbf{Main Research Use} 
& \textbf{Data Types} 
& \textbf{Real / Synthetic} 
& \textbf{Data Size (Number of Records)} \tabularnewline
\hline

YCB Dataset \cite{calli2015ycb}
& Object detection, robotic manipulation, grasping
& RGB images, 3D models, object meshes
& Real
& 77 objects, 92 different object textures \tabularnewline
\hline

GraspNet \cite{fang2020graspnet}
& Robotic grasping, object manipulation 
& RGB-D images, camera poses, ground truth 6D object poses
& Real 
& 97,280 objects, 1.1M grasp annotations \tabularnewline
\hline

APC (Amazon Picking Challenge) \cite{zeng2017multi} 
& Object picking, robotic shelving, manipulation
& RGB images, depth maps, object masks
& Real 
& 280,000 images for 39 objects \tabularnewline
\hline

OmniObjects \cite{bousmalis2018using}
& Simulated to real transfer, robotic grasping, manipulation
& RGB images, depth maps, 3D object models
& Synthetic + Real
& 15,000 simulated grasp scenes, 400 real object scenes \tabularnewline
\hline

Toybox Dataset \cite{wang2018toybox}
& Object interaction, robotic manipulation, object recognition
& RGB videos with varied viewpoints, interactions
& Real 
& 3+ hours video per category, 12 object categories \tabularnewline
\hline

Jacquard Dataset \cite{depierre2018jacquard} 
& Robotic grasping, object manipulation planning, grasp synthesis
& RGB-D images, object segmentation, grasp labels
& Synthetic 
& 11,000 objects, over 54,000 grasp attempts \tabularnewline
\hline

RobotriX Challenge \cite{zhang2021robotrix}
& Object tracking, manipulation, multi-agent tasks
& RGB-D images, simulated environment scenes
& Synthetic + Real
& Over 350 robot manipulation tasks \tabularnewline
\hline

ROB811 - Robotic Manipulation Dataset \cite{rob811}
& General robotic manipulation, object interactions 
& RGB-D sensor data, object poses 
& Real
& Multiple manipulation tasks with over 100 object instances \tabularnewline
\hline

MetaWorld (Reinforcement Learning) \cite{yu2020meta}
& Robotic control, reinforcement learning, meta-learning
& Simulated robotic control, robotic arm manipulation
& Synthetic
& 50+ robotic manipulation tasks \tabularnewline
\hline

\end{tabular}
}
\end{table*}

\section{Challenges and Failure Modes}\label{sec:fail_ai_core}

\subsection{AI Core Algorithm}\label{sec:ai_core_alg}

\begin{table*}[htbp]
\centering
\caption{Overview of Large Language Models (LLMs) and Large Vision-Language Models (LVLMs).}
\label{tab:llms_and_lvms}
\footnotesize
\resizebox{\textwidth}{!}{
\begin{tabular}{lccccccc}
\toprule
\textbf{Model Family} & \textbf{Model Name} & \textbf{Size (Billion)} & \textbf{Base Model} & \textbf{Pre-train Data} & \textbf{Hardware} & \textbf{Training Time} & \textbf{Release Date} \\
\midrule

\multicolumn{8}{c}{\textbf{Large Language Models (LLMs)}} \\
\midrule

\multirow{7}{*}{\textbf{GPT Family}} 
& GPT-1 & 0.12B & - & 1.3 Billion Tokens & - & - & 2018-06 \\
& GPT-2 & 1.5B & - & 10 Billion Tokens & - & - & 2019-02 \\
& GPT-3 & 125–175B & - & 0.3 Trillion Tokens & - & - & 2020-05 \\
& Codex & 12B & GPT-3 & 100 Billion Tokens & - & - & 2021-07 \\
& WebGPT & 0.76–175B & GPT-3 & - & - & - & 2021-12 \\
& InstructGPT & 175B & GPT-3 & - & - & - & 2022-03 \\
& GPT-4 & - & - & - & - & - & 2023-03 \\

\midrule
\multirow{7}{*}{\textbf{LLaMA Family}} 
& LLaMA & 7–65B & - & 1.4 Trillion Tokens & 2048 A100 & 21 days & 2023-02 \\
& LLaMA2 & 7–70B & - & 2 Trillion Tokens & 2000 A100 & - & 2023-07 \\
& Alpaca & 7B & LLaMA1 & - & - & - & 2023-03 \\
& Mistral-7B & 7.3B & - & - & - & - & 2023-09 \\
& Code Llama & 34B & LLaMA2 & 500 Billion Tokens & - & - & 2023-08 \\
& LongLLaMA & 3B, 7B & OpenLLaMA & 1 Trillion Tokens & - & - & 2023-10 \\
& LLaMA-Pro-8B & 8.3B & LLaMA2-7B & 80 Billion Tokens & - & - & 2023-11 \\

\midrule
\multirow{7}{*}{\textbf{PaLM Family}} 
& PaLM & 8–540B & - & 0.78 Trillion Tokens & 6144 TPU v4 & - & 2022-04 \\
& U-PaLM & 8–540B & - & 1.3 Billion Tokens & - & - & 2022-06 \\
& Flan-PaLM & 540B & PaLM & - & 512 TPU v4 & 37 h & 2022-10 \\
& Flan-U-PaLM & 540B & U-PaLM & - & - & - & 2022-10 \\
& Med-PaLM & 540B & PaLM & 0.78 Trillion Tokens & - & - & 2022-10 \\
& PaLM-2 & 340B & - & 3.6 Trillion Tokens & - & - & 2023-05 \\
& Med-PaLM 2 & - & PaLM-2 & - & - & - & 2023-05 \\

\midrule
\multirow{8}{*}{\textbf{Other Models}} 
& GShard & 600B & - & 1 Trillion Tokens & 2048 TPU v3 & 4 days & 2020-06 \\
& HyperCLOVA & 82B & - & 300 Billion Tokens & 1024 A100 & 13.4 days & 2021-09 \\
& FLAN & 137B & LaMDA-PT & - & 128 TPU v3 & 60 h & 2021-09 \\
& Yuan 1.0 & 245B & - & 180 Billion Tokens & 2128 GPU & - & 2021-10 \\
& Anthropic & 52B & - & 400 Billion Tokens & - & - & 2021-12 \\
& Gopher & 280B & - & 300 Billion Tokens & 4096 TPU v3 & 920 h & 2021-12 \\
& BLOOM & 176B & - & 0.366 Trillion Tokens & 384 A100 & 105 days & 2022-11 \\

\midrule
\multicolumn{8}{c}{\textbf{Large Vision-Language Models (LVLMs)}} \\
\midrule

\multirow{6}{*}{\textbf{CLIP Family}} 
& CLIP & 0.4B & ResNet/Vit-B & 400M Image-Text Pairs & - & - & 2021-01 \\
& OpenCLIP & 0.4–1.0B & ViT-L/14 & LAION-400M & - & - & 2021-12 \\
& CLIP-ViT-G & 1.8B & ViT-G & LAION-2B & - & - & 2022-10 \\
& CLIP-ViT-H & 2.0B & ViT-H & LAION-5B & - & - & 2023-01 \\
& CLIP-ViT-Large & 1.0B & ViT-L & LAION-5B & - & - & 2023-03 \\
& CLIP-ViT-XL & 2.5B & ViT-XL & LAION-5B & - & - & 2023-06 \\

\midrule
\multirow{6}{*}{\textbf{ALIGN Family}} 
& ALIGN & 1.8B & EfficientNet-L2 & 1.8B Image-Text Pairs & TPU v4 & - & 2021-05 \\
& CoCa & 2.1B & ViT-G & JFT-4B + Image-Text Pairs & TPU v4 & - & 2022-06 \\
& Flamingo & 80B & Chinchilla & 2.3B Image-Text Pairs & TPU v4 & - & 2022-04 \\
& PaLI & 17B & ViT-e & 10B Image-Text Pairs & TPU v4 & - & 2022-10 \\
& PaLI-X & 55B & ViT-e & 27B Image-Text Pairs & TPU v4 & - & 2023-05 \\
& Gemini & 175B & Gemini-L & Multimodal Dataset & TPU v5 & - & 2023-09 \\

\midrule
\multirow{6}{*}{\textbf{BLIP Family}} 
& BLIP & 0.4B & ViT-B & 129M Image-Text Pairs & - & - & 2022-01 \\
& BLIP-2 & 0.4B & ViT-L & 129M Image-Text Pairs & - & - & 2023-02 \\
& MiniGPT-4 & 7B & Vicuna & 3M Image-Text Pairs & - & - & 2023-04 \\
& LLaVA & 13B & LLaMA & 595K Image-Text Pairs & - & - & 2023-05 \\
& Otter & 13B & LLaMA & 2M Image-Text Pairs & - & - & 2023-06 \\
& InstructBLIP & 13B & LLaMA & 1.2M Image-Text Pairs & - & - & 2023-07 \\

\midrule
\multirow{6}{*}{\textbf{Other Models}} 
& Kosmos-1 & 1.6B & Transformer & Multimodal Dataset & - & - & 2023-03 \\
& Kosmos-2 & 2.0B & Transformer & Multimodal Dataset & - & - & 2023-08 \\
& GIT & 0.8B & Swin Transformer & 800M Image-Text Pairs & - & - & 2022-06 \\
& GIT-2 & 1.0B & Swin Transformer & 1.2B Image-Text Pairs & - & - & 2023-01 \\
& SEED & 1.0B & ViT-L & 1.1B Image-Text Pairs & - & - & 2023-04 \\

\bottomrule
\end{tabular}}
\end{table*}

\begin{figure*}
    \centering
    \includegraphics[width=\linewidth]{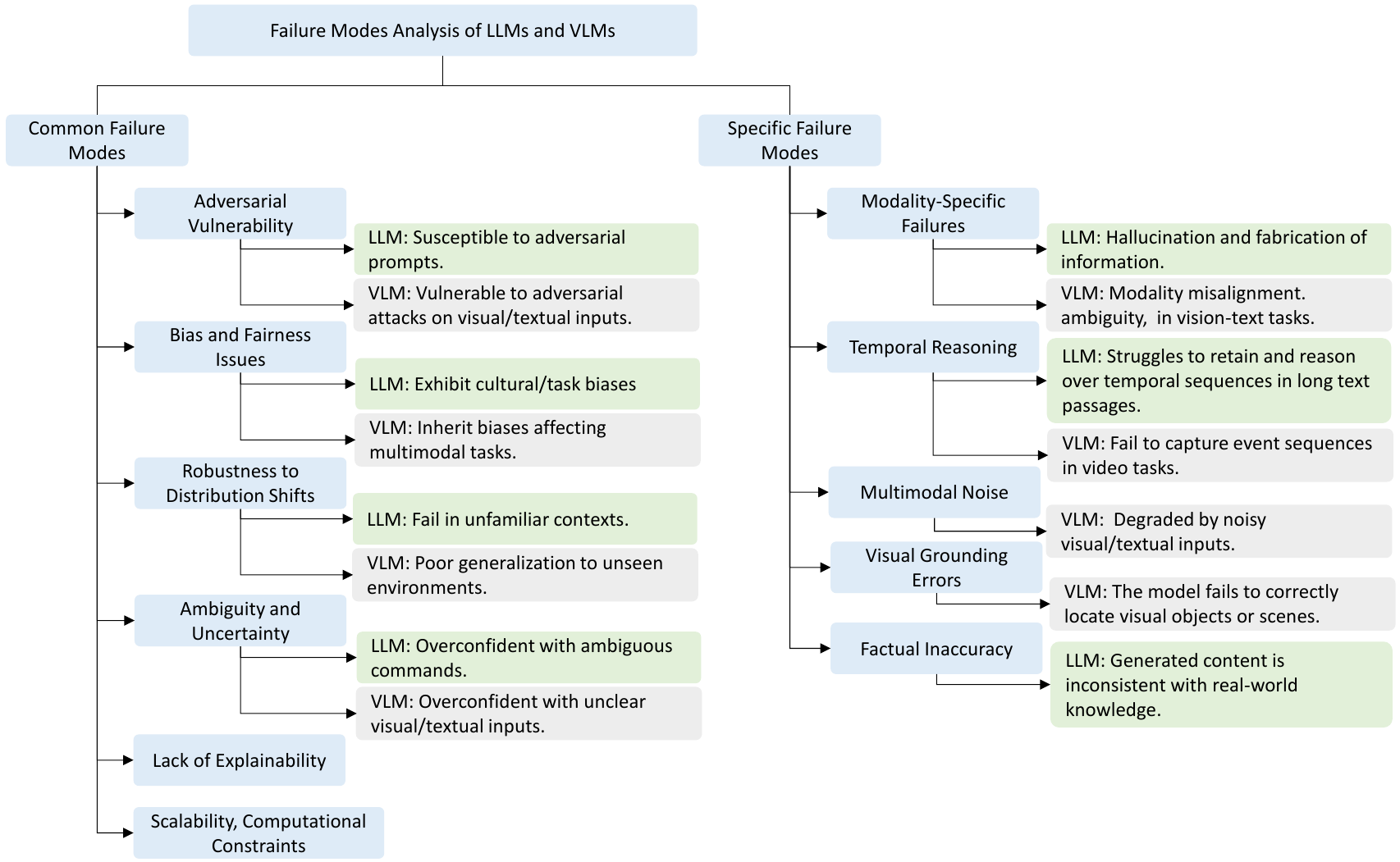}
    \caption{Failure modes classification.}
    \label{fig:enter-label}
\end{figure*}

\subsubsection{Large Vision Language Models}\label{sec:vlm_lmm}

Embodied AI systems powered by LVLMs are revolutionizing robotics by enabling machines to visually perceive, understand, and interact with their environments. These models enhance contextual awareness and improve task execution through advanced multimodal integration and reasoning.

LVLMs, such as CLIP, BLIP-2, and GPT-4 Vision, excel in tasks requiring cohesive visual-linguistic reasoning, while models like OFA and PaLI provide versatility by addressing diverse vision-language tasks within a unified framework. Real-time applications benefit from low-latency models like mPLUG and LLaVA, optimized for dynamic environments. Ethical considerations are also prioritized in closed-source models like Claude 3 Vision and PaLM-E, which focus on safety and alignment. Additionally, robust models such as BLIP and Flamingo handle noisy or incomplete data effectively, ensuring reliability in real-world scenarios.

LMMs further extend these capabilities by processing and reasoning across diverse data modalities, including images, text, and multilingual content. For instance, Wang \cite{wang2024large} proposed a GPT-4V-based framework that integrates natural language instructions with robotic visual perception to generate detailed, real-world-aligned action plans, validated on multiple robotic datasets. Similarly, ViLA \cite{gruber2008look} leverages GPT-4V for closed-loop adaptability, dynamically refining action plans based on visual feedback to handle environmental changes and execute long-term tasks effectively. MultiPLY \cite{hong2024multiply}, built on LLaVA \cite{liu2024visual}, adopts an object-centric approach using action tokens (e.g., navigating, picking up) and state tokens (e.g., tactile feedback, point cloud data) to guide embodied agents in interacting with their surroundings. By transitioning seamlessly between abstract reasoning and detailed multimodal observations, MultiPLY demonstrates versatility across diverse interactive scenarios.

While these advancements highlight the transformative potential of LVLMs and LMMs in enabling robust, adaptive, and context-aware embodied AI, challenges remain. LVLMs, despite their strengths in tasks like image captioning, visual question answering (VQA), and multimodal reasoning, exhibit limitations in real-world applications. These include failure modes that compromise performance, safety, and reliability, particularly in dynamic and complex environments. Addressing these limitations is critical to fully realizing the potential of embodied AI systems.

\subsubsection{Large Language Models}\label{sec:llms}

LLMs have become a cornerstone of modern artificial intelligence, excelling in diverse natural language processing tasks such as machine translation, summarization, question answering, and content generation \cite{radford2019language,devlin2018bert,brown2020language,openai2023gpt4,schulman2023introducing,park2023generative,west2023advances}. Their integration into embodied AI systems, where agents interact with the physical world, has further expanded their role, with LLMs often serving as the cognitive core. In this capacity, they process multimodal inputs—such as language commands, visual data, and sensor readings—and generate contextually appropriate responses or actions to guide the agent's behavior. However, despite their remarkable capabilities, deploying LLMs in embodied systems brings significant challenges. A particularly urgent concern is their vulnerability to adversarial attacks, including black-box attacks, where an adversary manipulates the system's behavior without direct access to the model's internal parameters. Such attacks exploit inherent weaknesses, such as susceptibility to adversarial prompts or reliance on incomplete or biased training data. These vulnerabilities underscore the critical need for research aimed at enhancing the safety, robustness, and reliability of LLM-driven embodied systems, especially as they increasingly interact with complex and dynamic real-world environments.

\subsection{Common Failure Modes}\label{sec:cfm}
LLMs and LVLMs exhibit several overlapping failure modes, while also facing unique challenges stemming from the specific modalities they process. This section focuses on the shared failure modes that arise across both types of models, providing a foundation for understanding their broader limitations and vulnerabilities.

\paragraph{Adversarial Vulnerability}\label{sec:adv_vul}
Both LLMs and LVLMs are susceptible to adversarial attacks, though the specifics vary across modalities. LLMs are particularly vulnerable to adversarial prompts, where carefully crafted malicious or ambiguous inputs can elicit harmful, misleading, or unsafe outputs. Similarly, LVLMs face adversarial threats in both the visual and textual domains; small, often imperceptible perturbations in input images or text can result in significant errors, such as misclassifying objects or generating inaccurate captions. A shared challenge between the two lies in their susceptibility to adversarial manipulation, which can compromise their performance and potentially lead to unsafe or incorrect actions when deployed in embodied AI systems.

\paragraph{Bias and Fairness Issues}\label{sec:bias_fair}
Both LLMs and LVLMs are prone to biases inherited from their training data, which can manifest in various ways and lead to unfair or harmful outcomes. LLMs, for instance, may exhibit biases in human-robot interactions, such as prioritizing certain tasks unfairly or misinterpreting cultural nuances, which can hinder effective communication or collaboration. Similarly, LVLMs often reflect biases present in their training datasets, resulting in skewed outputs in tasks like image captioning or multimodal reasoning. A common challenge for both models is their tendency to perpetuate these biases, which can be particularly problematic in diverse or sensitive environments, where fairness and inclusivity are critical.

\paragraph{Robustness to Distribution Shifts}\label{sec:robust_dist_shift}
LLMs and LVLMs both face significant challenges in generalizing to environments or inputs that differ from their training data, which can result in failures and unsafe behavior. For LLMs, this often manifests as difficulty in handling unfamiliar environments or domain-specific language, leading to misinterpretation of commands or poor decision-making. Similarly, LVLMs struggle to generalize effectively in dynamic or diverse real-world settings, where inputs may deviate significantly from the data they were trained on. A shared limitation of these models is their reliance on training data, which constrains their ability to adapt to novel or unpredictable scenarios, ultimately impacting their reliability and safety in real-world applications.

\paragraph{Ambiguity and Uncertainty}\label{sec:amg_uncert}
Both LLMs and LVLMs exhibit significant limitations in dealing with ambiguity and uncertainty, often leading to overconfident outputs that can compromise safety and reliability in embodied systems. LLMs frequently generate overconfident responses to vague or ambiguous commands, which can result in unsafe or unintended actions, particularly in high-stakes scenarios. Similarly, LVLMs struggle with incomplete or ambiguous inputs, producing predictions with high confidence even when the visual or textual data lacks clarity or context. A shared challenge for these models is their inability to effectively express or quantify uncertainty, which is critical for ensuring safe and cautious behavior in real-world applications. This overconfidence amplifies the risk of errors, especially in environments where ambiguity is common or unavoidable.

\paragraph{Lack of Explainability}\label{sec:lack_expl}
A significant challenge shared by LLMs and LVLMs is their lack of explainability, which poses serious risks in safety-critical applications such as healthcare, autonomous systems, or robotics. LLMs function as "black boxes," making it difficult to trace or understand the reasoning behind their outputs, which complicates error diagnosis and debugging. This opacity becomes particularly problematic when these systems are tasked with making decisions that directly impact human safety or well-being. Similarly, LVLMs exhibit a lack of transparency in their decision-making processes, making it challenging to interpret their predictions or identify the root causes of failures. The shared absence of explainability in both models undermines trust, accountability, and the ability to improve their performance, particularly in sensitive or high-stakes environments where understanding system behavior is essential for ensuring reliability and safety.

\paragraph{Scalability and Computational Constraints}\label{sec:scal_comp_constraint}
LLMs and LVLMs both face significant challenges due to their high computational demands, which impact their scalability and real-time performance in practical applications. LLMs require substantial computational resources, often resulting in latency or even failures when operating in real-time, resource-constrained environments. Similarly, LVLMs encounter scalability issues, as their large-scale architectures make them difficult to deploy on devices with limited processing power or in systems that demand real-time responsiveness. Both types of models share the common limitation of being computationally expensive, which restricts their integration into embodied AI systems that require efficiency, scalability, and reliable performance under resource constraints.

\subsection{Algorithm Specific Failure Modes}\label{sec:algorithm_specif_fail}

\paragraph{Modality-Specific Failures}\label{sec:modal_specific_failure}
    \begin{itemize}
        \item \textbf{LLMs}: LLMs excel at processing and generating text, making them valuable for reasoning and planning in robotic tasks, but their limitations pose significant challenges in embodied AI systems. A key issue is their lack of grounding in the physical world, as they struggle to interpret or act upon real-world sensory data. This disconnect between language and physical reality often results in failure modes such as hallucination or fabrication of information, where LLMs generate plausible-sounding but incorrect or misleading instructions. Such errors can lead to unsafe actions, such as misidentifying objects or providing faulty navigation guidance. Additionally, LLMs inherit biases from their large-scale training datasets, perpetuating harmful stereotypes \cite{sheng2019woman,groenwold2020investigating,abid2021persistent,venkit2022study,nadeem2020stereoset,blodgett2021stereotyping}, which can manifest in discriminatory or unethical decision-making. They are also prone to generating misinformation \cite{lin2021truthfulqa,weidinger2022taxonomy,menick2022teaching,buchanan2021truth}, which can be especially dangerous in high-stakes applications like healthcare or autonomous systems, where incorrect outputs may cause physical harm or system failures. Despite efforts to filter harmful content, LLMs can produce toxic or offensive language, particularly when exposed to adversarial prompts \cite{gehman2020realtoxicityprompts}, leading to inappropriate interactions in embodied settings. Privacy violations are another concern, as LLMs may inadvertently memorize and reproduce sensitive information, compromising user data or operational confidentiality \cite{carlini2021extracting}. Furthermore, they are vulnerable to psychological manipulation, where adversarial inputs can exploit system vulnerabilities or deceive users in conversational interactions \cite{roose2023conversation,elatillah2023man}. While external affordance models, such as open-vocabulary detectors \cite{minderer2022simple} and value functions \cite{ahn2022can}, have been proposed to ground LLMs in the physical world \cite{gibson1977theory}, these approaches often fall short in complex environments. Such models typically function as one-way channels, lacking the ability to effectively convey task-specific information \cite{gruber2008look}, further limiting the integration of LLMs into embodied AI systems.

        \item \textbf{LVLMs}: LVLMs, which handle both visual and textual inputs, face significant challenges related to modality misalignment, where the model fails to correctly associate visual features with corresponding text. This misalignment can lead to two primary issues: (1) \textit{Incorrect Image-Text Matching}, where the model may erroneously pair an image with an unrelated text description, particularly in tasks like zero-shot image classification, where subtle or ambiguous visual cues are misinterpreted; and (2) \textit{Failure in Visual Grounding}, where the model struggles to link language to specific visual regions, such as identifying relevant objects in tasks like object detection or referring expression comprehension. This is especially problematic in embodied AI systems, where precise object identification is critical for task execution. Compared to LLMs, which primarily generate incorrect information in linguistic contexts, LVLMs uniquely struggle with the integration and alignment of visual and textual modalities.
    \end{itemize}

\paragraph{Temporal Reasoning}\label{sec:temporal_reason}
    \begin{itemize}
\item \textbf{LVLMs}: LVLMs face significant challenges in temporal reasoning, particularly in video-based tasks that require understanding event sequences and long-term dependencies. These models often struggle with \textit{temporal inconsistencies}, such as misinterpreting the order of actions or failing to recognize changes in an object's state over time, as well as difficulties in maintaining \textit{long-term dependencies}, which can result in incomplete predictions. For example, in a cooking video, a VLM might fail to link earlier actions like chopping vegetables to later steps like adding them to a pot, leading to incorrect interpretations of dynamic scenes.
\item \textbf{LLMs}: While LLMs can handle sequential data (e.g., text), they do not typically face the same challenges with temporal reasoning in visual contexts.
    \end{itemize}

\paragraph{Adversarial Multimodal Noise}\label{sec:adv_MM_nois}
    \begin{itemize}
        \item \textbf{LVLMs}: Struggle with \textit{multimodal noise}, where noisy or incomplete visual and textual inputs degrade performance, especially in real-world environments with poor lighting, occlusions, or motion blur. They are highly vulnerable to adversarial attacks. Small, imperceptible changes in visual inputs, such as adding noise to an image, can lead to misclassifications (e.g., labeling a cat as a dog), while subtle modifications to textual prompts can result in incorrect or harmful outputs, such as providing the wrong answer in a visual question answering task. These vulnerabilities pose significant risks, particularly in embodied AI systems
        \item \textbf{LLMs}: While LLMs can also struggle with noisy textual inputs, they do not face the same challenges with noisy visual data. LVLMs must contend with noise in both visual and textual modalities, whereas LLMs primarily deal with noise in language inputs.
    \end{itemize}

\begin{table*}[t]
\centering
\caption{Comparison of Evaluation Datasets: General, Adversarial, and Safety Evaluation Tasks}
\label{tab:lvml_datasets}
\resizebox{\textwidth}{!}{
    \centering
\begin{tabular}{llcccc}
\toprule
\textbf{Category} & \textbf{Dataset Name} & \textbf{\# Tasks} & \textbf{Sample Type} & \textbf{\# Samples} & \textbf{Features} \\
\midrule
\multirow{10}{*}{{General Datasets}} 
    & ImageNet \cite{deng2009imagenet}          & 1    & Image              & 14M           & General \\
    & RefCOCOg \cite{p7}          & 1    & Image-Text Pair    & 85k           & General \\
    & RefCOCO+ \cite{p6}          & 1    & Image-Text Pair    & 141k          & General \\
    & RefCOCO \cite{p5}           & 1    & Image-Text Pair    & 142k          & General \\
    & COCO Captions \cite{p4}     & 1    & Image-Text Pair    & 1M            & General \\
    & Flickr30k \cite{p8}         & 1    & Image-Text Pair    & 159k          & General \\
    & OK-VQA \cite{p10}           & 1    & Image-Question Pair & 14k          & General \\
    & VQA V2 \cite{p9}            & 1    & Image-Question Pair & 1.1M         & General \\
    & LVLM-eHub \cite{p11}        & 47   & Image-Text Pair    & 333k          & General \\
    & Tiny LVLM-eHub \cite{p12}   & 42   & Image-Text Pair    & 2.1k          & General \\
\midrule
\multirow{12}{*}{{Adversarial Datasets}} 
    & RedTeam-2K \cite{p17}       & 16   & Text               & 2k            & Red Team \\
    & MultiJail \cite{p18}        & 18   & Text               & 3.15k         & Red Team \\
    & HarmBench \cite{p24}        & 7    & Text-Image Pair    & 510           & Red Team \\
    & Achilles \cite{p25}         & 5    & Image-Text Pair    & 750           & Red Team \\
    & MM-SafetyBench \cite{p30}   & 13   & Image-Text Pair    & 5.04k         & Attack Samples, Red Team \\
    & SALAD-Bench \cite{p31}      & 66   & Text               & 30k (21.3k)   & Attack Samples, Red Team \\
    & JBB-Behaviors \cite{p23}    & 10   & Text               & 100           & Red Team \\
    & AdvBench \cite{p21}         & 8    & Text-Image Pair    & 1k            & Red Team \\
    & SafeBench \cite{p20}        & 10   & Text               & 500           & Red Team \\
    & JailbreakHub \cite{p29}     & 13   & Text               & 100k (390)    & Attack Samples, Red Team \\
    & OOD-VQA \cite{p34}          & 8    & Image-Question Pair & 8.2k         & Adversarial Robustness \\
    & RTVLM \cite{p33}            & 10   & Image-Text Pair    & 5.2k          & Adversarial Robustness \\
    & AVIBench \cite{p32}         & 6    & Text-Image Pair    & 260k          & Adversarial Robustness \\
    & JailBreakV-28K \cite{p17}   & 16   & Text-Image Pair    & 28k           & Adversarial Robustness \\
\midrule
\multirow{10}{*}{{Alignment Datasets}} 
    & SafetyBench \cite{p35}      & 7    & Text-Question Pair & 11.4k         & Toxicity Sensitivity \\
    & RealToxicityPrompts \cite{p36} & 8 & Text               & 100k          & Toxicity Sensitivity \\
    & BeaverTails \cite{p42}      & 14   & Text               & 30k           & Toxicity Training \\
    & SPA-VL \cite{p43}           & 53   & Image-Question Pair & 100k         & Toxicity Training \\
    & ToxicChat \cite{p39}        & 2    & Text               & 10k           & Toxicity Training \\
    & hh-rlhf \cite{p40, p41}     & 20   & Text               & 44k           & Toxicity Training \\
    & Safety-Prompts \cite{p38}   & 8    & Text               & 100k          & Toxicity Training \\
    & ToViLaG \cite{p37}          & 3    & Text               & 33k           & Toxicity Sensitivity \\
\bottomrule
\end{tabular}}
\end{table*}

\section{Dataset Taxonomy for LLMs and LVLMs Evaluation}\label{sec:dataset_taxono_llm_vlm}

To comprehensively evaluate large LVLMs, datasets can be categorized into four key types: General Datasets, Red Team Datasets, Robustness Evaluation Datasets, and Alignment Datasets. Each serves a distinct purpose in assessing various aspects of model performance and robustness.

\paragraph{General Datasets}\label{sec:general_dataset}
General datasets focus on evaluating core multimodal capabilities, such as image classification, captioning, visual question answering (VQA), testing models' visual understanding, reasoning, and language generation. Popular benchmarks include ImageNet, COCO Captions, RefCOCO, and VQA V2. These datasets can also be adapted for robustness testing by introducing subtle perturbations to simulate cognitive biases.

\paragraph{Adversarial Datasets}\label{sec:adv_dataset}
Adversarial datasets are specialized for stress-testing models and are divided into two main types:

\begin{itemize}
     \item Red Team Datasets: These datasets target overtly harmful content, such as violence, explicit material, or other policy-violating inputs. They are used to evaluate model robustness against malicious queries, ensure ethical compliance, and simulate jailbreak scenarios. Examples include RedTeam-2K \cite{luo2024jailbreakv}, MultiJail \cite{deng2023multilingual}, and SALAD-Bench \cite{li2024salad}.
\item Robustness Evaluation Datasets assess model resilience to adversarial attacks, ambiguous queries, and edge cases by exposing vulnerabilities in handling subtle or adversarial inputs. Subcategories include adversarial attack samples and sensitivity tests for harmful or misinterpreted inputs. For example, AVIBench \cite{zhang2024bavibenchevaluatingrobustnesslarge} generates diverse adversarial visual instructions to evaluate LVLM robustness, while RoCOCO \cite{park2024rococorobustnessbenchmarkmscoco} introduces adversarial text and images to test image-text matching models. Studies show that many state-of-the-art models suffer significant performance degradation when encountering such adversarial samples.

\end{itemize}

\paragraph{Alignment Datasets}\label{sec:align_dataset}
Alignment datasets are crucial for fine-tuning LVLMs, ensuring they balance helpfulness with harmlessness. Often used in RLHF pipelines or for preference model training, these datasets align models with ethical standards, minimizing harmful outputs while maintaining utility.
These datasets can be categorized into two main types:
\begin{itemize}
\item Preference-Based Alignment Datasets support preference modeling in RLHF and DPO, helping LVLMs balance safety and usability. For example, SPA-VL \cite{zhang2024spa} provides safety preference data across domains, aiding LVLMs in aligning responses with ethical guidelines.
\item Instruction-Based Alignment Datasets enhance instruction tuning to improve model alignment and reduce harmful outputs. VLFeedback \cite{li2024vlfeedback}, with over 82,000 multimodal instructions and AI-generated rationales, serves as a large-scale resource for vision-language alignment research.
\end{itemize}

\section{Conclusion}\label{sec:conclud}

The reliance of Embodied AI systems on intricate interactions between sensors, actuators, and algorithms exposes them to a wide range of vulnerabilities and security threats. This survey provides a comprehensive overview of the vulnerabilities and attack vectors targeting Embodied AI systems, with a particular focus on the unique challenges posed by the integration of LVLMs and LLMs.
We have categorized the vulnerabilities of Embodied AI into three main dimensions: Exogenous Vulnerabilities, Endogenous Vulnerabilities, and Inter-Dimensional Vulnerabilities. Exogenous vulnerabilities arise from external factors such as dynamic environments, physical attacks, adversarial attacks, and cybersecurity threats. Endogenous vulnerabilities stem from internal system failures, including hardware malfunctions, software bugs, and design flaws. Inter-Dimensional vulnerabilities occur at the intersection of external and internal factors, exacerbating system fragility.
In the context of adversarial attacks, we have explored various attack paradigms targeting Embodied AI systems, including sensor spoofing, adversarial attacks on perception systems, and attacks on LVLMs and LLMs. These attacks exploit the multimodal nature of these systems, introducing perturbations in visual, textual, or multimodal inputs to deceive models and disrupt system operations. We have also discussed the challenges in evaluating the robustness of these models, highlighting the need for comprehensive benchmarking and evaluation frameworks.
To address these challenges, we have proposed targeted strategies to enhance the safety and reliability of Embodied AI systems. These strategies include improving world grounding through voxel-based representations and neural radiance fields, enhancing multimodal integration, and implementing robust control and adaptation mechanisms. Additionally, we have emphasized the importance of safety-critical design and verification, including formal verification, Sim2Real testing, and redundant safety mechanisms.
In conclusion, this survey provides a comprehensive framework for understanding the interplay between vulnerabilities and safety in Embodied AI systems.

\newpage 
\bibliographystyle{abbrv}
\bibliography{reference}

\vfill

\end{document}


\title{Appendix}

\begin{table*}[ht]
\centering
\caption{Tesla Autopilot Failure Causes, Associated CWE, and Keywords (with Da Vinci Surgical Robot)}
\begin{tabular}{p{1.5cm}|p{2.5cm}|p{3.5cm}|p{8cm}}
\toprule
{Category} & {Cause} & {Description} & {Associated CWE and Keywords} \\
\midrule
\multirow{6}{1.5cm}{Autonomous Driving System} & {Sensor Misjudgment} & Failure to detect objects or handle adverse environmental conditions & CWE-20 (Improper Input Validation), CWE-824 (Access of Uninitialized Pointer), CWE-754 (Improper Check for Unusual or Exceptional Conditions) \\
\cline{2-4}
 & {Driver Over-Reliance on Autopilot} & Drivers fail to take over or are distracted & CWE-841 (Improper Enforcement of Behavioral Workflow), CWE-285 (Improper Authorization) \\
\cline{2-4}
 & {System Boundary Conditions} & Autopilot struggles in complex environments or non-highway scenarios & CWE-693 (Protection Mechanism Failure), CWE-754 (Improper Check for Unusual or Exceptional Conditions) \\
\cline{2-4}
& {Lane Keeping and Auto Lane Change Issues} & Lane departure or incorrect lane change decisions & CWE-1194 (Improper Control of Dynamically-Managed Code Resources), CWE-754 (Improper Check for Unusual or Exceptional Conditions) \\
\cline{2-4}
 & {Emergency Braking and Collision Avoidance Failures} & Failure to trigger emergency braking or avoid collisions & CWE-754 (Improper Check for Unusual or Exceptional Conditions), CWE-20 (Improper Input Validation) \\
\cline{2-4}
 & {Road Sign and Traffic Signal Recognition Issues} & Misinterpretation of traffic signals or confusion with road signs & CWE-20 (Improper Input Validation), CWE-1194 (Improper Control of Dynamically-Managed Code Resources) \\
\midrule
\multirow{5}{1.5cm}{Surgical Robot System} & {Robotic Arm Malfunction} & Mechanical failure of robotic arms during surgery & CWE-841 (Improper Enforcement of Behavioral Workflow), CWE-693 (Protection Mechanism Failure) \\
\cline{2-4}
 & {Console Malfunction} & Surgeon console errors, preventing proper control of the robot & CWE-841 (Improper Enforcement of Behavioral Workflow), CWE-693 (Protection Mechanism Failure) \\
\cline{2-4}
 & {Optic System Malfunction} & Failure in the camera or optical system, leading to loss of visual feedback & CWE-400 (Uncontrolled Resource Consumption), CWE-693 (Protection Mechanism Failure) \\
\cline{2-4}
 & {System Errors} & General system errors, including software crashes or unexpected shutdowns & CWE-754 (Improper Check for Unusual or Exceptional Conditions), CWE-693 (Protection Mechanism Failure) \\
\bottomrule
\end{tabular}
\end{table*}

\bibliographystyle{IEEEtran}
\bibliography{reference,main}

\vfill